\documentclass[fleqn,usenatbib,useAMS]{mnras}

\usepackage{graphicx}	
\usepackage{amsmath}	
\usepackage{amssymb}	
\usepackage{multicol}        
\usepackage{bm}		
\usepackage{pdflscape}	
\usepackage{ulem}

\newcommand{\msun}{M_{\odot}}

\newcommand{\mpeak}{M_{\rm peak}}
\newcommand{\mzero}{M_{0}}
\newcommand{\lgmzero}{m_{0}}
\newcommand{\mhalo}{M_{\rm halo}}
\newcommand{\rvir}{R_{\rm vir}}
\newcommand{\tform}{t_{\rm form}}

\newcommand{\cbl}{\beta_{\rm late}}
\newcommand{\tauc}{\tau_{\rm c}}
\newcommand{\pc}{\text{p}_{50\%}}
\newcommand{\cmin}{c_{\rm min}}
\newcommand{\cminprime}{\tilde{c}_{\rm min}}
\newcommand{\taucprime}{\tilde{\tau}_{\rm c}}
\newcommand{\cblprime}{\tilde{\beta}_{\rm late}}

\newcommand{\lgc}{\log_{10}c}
\newcommand{\dd}{d}

\usepackage[dvipsnames]{xcolor}
\definecolor{aphcommentcolor}{rgb}{1.0, 0., 0.}

\newcommand{\tauh}{\tau_{\rm h}}
\newcommand{\aearly}{\alpha_{\rm early}}
\newcommand{\alate}{\alpha_{\rm late}}
\newcommand{\thetanfw}{\theta_{\rm NFW}}

\usepackage{xspace}
\newcommand{\dprof}{{Diffprof}\xspace}
\newcommand{\dprofpop}{{DiffprofPop}\xspace}
\newcommand{\dmah}{{Diffmah}\xspace}
\newcommand{\dmahpop}{{DiffmahPop}\xspace}
\newcommand{\dstar}{{Diffstar}\xspace}
\newcommand{\dstarpop}{{DiffstarPop}\xspace}

\newcommand{\tperc}[1]{t_{#1\%}}

\newcommand{\beq}{\begin{eqnarray}}
\newcommand{\eeq}{\end{eqnarray}}
\newcommand{\ben}{\begin{enumerate}}
\newcommand{\een}{\end{enumerate}}
\newcommand{\bit}{\begin{itemize}}
\newcommand{\eit}{\end{itemize}}

\newcommand{\lgg}{\log_{10}}

\usepackage[T1]{fontenc}
\usepackage{ae,aecompl}
\usepackage{newtxtext,newtxmath}

\title[A Differentiable Model of Halo Concentration]{A Differentiable Model of the Evolution of Dark Matter Halo Concentration}

\author[Stevanovich, Hearin, \& Nagai]{Dash Stevanovich$^{1}$\thanks{Contact e-mail: \href{mailto:dash.stevanovich@yale.edu}{dash.stevanovich@yale.edu}}, Andrew P. Hearin$^{2}$, Daisuke Nagai$^{1}$
\\
$^{1}$ Department of Physics, Yale University, PO Box 208101, New Haven, CT, 06520, USA, and\\
$^{2}$ HEP Division, Argonne National Laboratory, 9700 South Cass Avenue, Lemont, IL, 60439, USA
}
\date{}

\begin{document}
\label{firstpage}
\pagerange{\pageref{firstpage}--\pageref{lastpage}}
\maketitle

\begin{abstract}
We introduce a new model of the evolution of the concentration of dark matter halos, ${\rm c}(t).$ For individual halos, our model approximates ${\rm c}(t)$ as a power law with a time-dependent index, such that at early times, concentration has a nearly constant value of ${\rm c}\approx3-4,$ and as cosmic time progresses, ${\rm c}(t)$ smoothly increases. Using large samples of halo merger trees taken from the Bolshoi-P and MDPL2 cosmological simulations, we demonstrate that our 3-parameter model can approximate the evolution of the concentration of individual halos with a typical  accuracy of 0.1 dex for $t\gtrsim2\,{\rm Gyr}$ for all Bolshoi-P and MDPL2 halos of present-day peak mass $\mzero\gtrsim10^{11.5}\msun.$ We additionally present a new model of the evolution of the concentration of halo populations, which we show faithfully reproduces both {\it average} concentration growth, as well as the {\it diversity} of smooth trajectories of ${\rm c}(t),$ including capturing correlations with halo mass and halo assembly history. Our publicly available source code, \href{https://github.com/BaryonPasters/diffprof}{\dprof}, can be used to generate Monte Carlo realizations of the concentration histories of cosmologically representative halo populations; \dprof is differentiable due to its implementation in the JAX autodiff library, which facilitates the incorporation of our model into existing analytical halo model frameworks.
\end{abstract}


\section{Introduction}
\label{sec:intro}

In the standard model of cosmology, the fundamental building blocks of structure formation are gravitationally self-bound halos of cold dark matter (CDM). These halos first form through the gravitational collapse of overdense patches of the initial density field, and then build up their mass over time via a combination of mergers and smooth accretion \citep[see][for a modern review]{mo_vdb_white_2010}. Observed galaxies are embedded in the centers of halos \citep{white_rees_1978,blumenthal_etal_1984}, and so the evolution of the internal structure of dark matter halos  plays an essential role in our theoretical understanding of the CDM framework for cosmological structure formation.

The radial density profile of dark matter halos is well-described by a double power law known as the NFW profile \citep{nfw_1997}, which is defined by an outer boundary, $R_{\rm halo},$ and by a single parameter encoding the internal structure, $c,$ the concentration of the halo, which exhibits a well-studied dependence upon total mass \citep[e.g.,][]{avila_reese_1999,dolag_etal04,diemer_kravtsov_2015,child_habib_etal_2018}. It has been known for many years that the concentration of a halo is fundamentally linked to its mass assembly history \citep{bullock_etal01, wechsler_etal02,vandenbosch02, zhao_etal03}. These early works established a relatively simple picture of dark matter halo growth that is comprised of two distinct phases. At early times, halo mass increases rapidly while concentration remains nearly constant with a value of $c\approx3-4.$ At later times, the rate of mass growth slows down, and halos tend to pile up mass onto their outskirts: during the slow-accretion phase, the central density of the halo remains relatively constant, while the concentration grows due to the increase of the halo boundary. This basic picture of halo structure growth has recently been confirmed in a model-independent fashion through an interpretable deep learning framework \citep{lucie_smith_etal23}. Even though mergers trigger large transient fluctuations in concentration \citep{wang_etal20,lucie_smith_etal22}, the internal structure of a halo has a remarkably durable memory, and even one-to-one mergers ultimately leave the central density of the halo undisturbed \citep{kazantzidis_etal06,vass_etal09,drakos_etal19}.

The connection between concentration and halo mass assembly plays an especially important role in efforts to derive cosmological constraints from astronomical measurements of massive halos. In the era of multi-wavelength cluster surveys, from microwave \citep{bocquet_etal19,aiola_etal20_act_cosmology}, optical \citep{abbott_etal20_des_y1_clusters}, to X-ray \citep{Vikhlinin_etal09} bands, one must rely upon an observational proxy that scales with halo mass, generically referred to as the ``mass--observable'' relation \citep[see][for a recent review]{Pratt2019ClusterMassReview}.
In order to realize the statistical power of upcoming surveys, each effort requires a detailed characterization of the associated mass--observable relation.

A successful model for the mass--observable relation needs to capture not only average trends, but also scatter that may exhibit physically important residual correlations \citep{stanek_etal10}. There are many indications that variations in the assembly history of massive halos are a major contributor to scatter in the ICM profiles \citep{lau_etal15}, cluster shapes \citep{chen_etal19,MachadoPolettiValle_etal21}, the Sunyaev-Zel'dovich (SZ) Effect \citep{yu_etal15,green_etal20}, and hydrostatic mass bias \citep{nelson_etal12,nelson_etal14,shi_etal15,shi_etal16}. On theoretical grounds, hydrodynamical simulations generically predict that scatter between different mass--observable relations should be correlated due to mutual covariance with cluster assembly \citep{wu_etal15_rhapsody,farahi_etal19}. There is also observational support for the notion that scatter in the mass--observable relation is driven by variations in internal structure and assembly history, such as a correlation between the scatter in the stellar mass of Brightest Cluster Galaxies (BCGs) and the halo mass and concentration of optical clusters \citep{zu_etal21,huang_etal22}. Mass assembly correlations may also be responsible for the reduction in scatter in optical estimations of cluster mass that can be achieved by leveraging the magnitude gap as an observational proxy for cluster formation history \citep{hearin_etal13,farahi_etal20}.

Numerous authors have capitalized upon the simplicity of the connection between halo growth and internal structure to build highly effective models for the evolution of halo concentration. In \citet{Zhao_2009}, the authors developed a model in which the behavior of $c(t)$ is determined by $\tperc{4},$ the time the halo first reached $4\%$ of its present-day mass. A remarkably successful simplification was first identified in \citet{ludlow_etal13}, in which it was found that the mean density of a halo interior to its scale radius, $\rho_{-2}\equiv\rho_{\rm NFW}(r_{-2}),$ is directly proportional to $\rho_{\rm crit}(t_{-2}),$ the critical density of the Universe evaluated at the time $t_{-2},$ where $t_{-2}$ is defined by $M_{\rm halo}(t_{-2})=M_{r<r_{-2}}(t_0),$ so that $t_{-2}$ is the time when the mass of the halo is equal to the amount of present-day mass enclosed within $r_{-2}.$ The simple picture suggested by this result is that halo concentration is roughly set by $\rho_{\rm crit}$ at the formation-time of the halo, $t_{\rm form},$ from which it follows naturally that halo concentration and $t_{\rm form}$ will be tightly correlated. This simplification was further leveraged in \citet{correa_etal15}, in which the authors used techniques from extended Press-Schechter theory \citep[e.g.,][]{zentner_eps_2007,jiang_vdb13} to build a model that faithfully captures median concentration, $\langle c\vert\mhalo, t\rangle_{\rm med},$ over a broad range of mass, time, and cosmological parameters.

In this paper, we develop a new model of the evolution of the concentration of individual and populations of dark matter halos. The basis of our approach is to parametrize {\it smooth trajectories} of $c(t),$ and then to characterize the statistical distribution of these trajectories across halo mass and redshift. We build upon the differentiable population modeling framework described in \citet{Hearin_2021}, which allows us to design our model to capture both the {\it average} growth in halo concentration, as well as the {\it diversity} observed in the concentration histories of simulated halos.

Using merger trees from the Bolshoi-P and MDPL2 N-body simulations described in \S\ref{sec:sims}, we approximate $c(t)$ for each simulated halo using the fitting function described in \S\ref{sec:individual_conc}.
In \S\ref{sec:conc_populations}, we describe our model of how the probability distribution of $c(t)$ depends upon $\mhalo$ and $\tform;$ throughout the paper, we will use the present-day {\it peak} halo mass, $\mzero,$ to characterize the mass-dependence of $c(t)$, as this choice anticipates our future applications of incorporating our results into a larger forward modeling pipeline. Here we demonstrate that for all Bolshoi-P and MDPL2 halos of present-day peak mass $\mzero\gtrsim10^{11.5}\msun,$ our model can accurately capture the {\it average} concentration growth, $\langle c(t)\vert\mzero\rangle,$ the {\it variance} in $P(c(t)\vert\mzero),$ as well as correlations between $c(t)$ and $\tform.$ We discuss our results and outline future applications in \S\ref{sec:discussion}, and conclude in \S\ref{sec:conclusion} with a summary of our principal findings. Mathematical and computational details underlying our results can be found in the appendices.


\section{Simulations and Merger Trees}
\label{sec:sims}
To construct our model for concentration histories, we use two gravity-only simulations: the Bolshoi-Planck simulation \citep[Bolshoi-P,][]{Klypin_2011}, and the MultiDark Planck 2 simulation \citep[MDPL2,][]{Klypin_2016}. The Bolshoi-P simulation evolves $2048^3$ dark matter particles with mass $m_{\rm p} = 1.55\times 10^8 M_{\sun}$ in a periodic box of width $250$ Mpc using the ART code \citep[][]{Kravtsov_1997} with a force resolution of 1 kpc. The MDPL2 simulation evolves $3840^3$ dark matter particles with mass $m_{\rm p}=1.51\times 10^9M_{\sun}$ in a box of width $1000$ Mpc using the L-GADGET-2 code \citep[][]{Springel_2005} with a force resolution of 5 kpc at low redshift and 13 kpc at high redshift. Cosmological parameters for these simulations are closely aligned with \citet[][]{Planck_2014}, and for both simulations we used the publicly available merger trees as identified by Rockstar and Consistent-Trees \citep[][]{Behroozi_2013a,Behroozi_2013b,rodriguez_puebla_etal16}. In these catalogs, the concentration of each halo was computed by dividing halo particles into up to 50 radial bins of equal mass (subject to the constraint of having at least 15 particles per radial bin), directly fitting the radial resulting density with an NFW functional form \citep{nfw_1997}, and selecting the maximum-likelihood of the fit. Further details about these simulations can also be found at the \href{https://www.cosmosim.org/}{CosmoSim} database \citep{riebe_etal13}.

\begin{figure*}
\includegraphics[width=17cm]{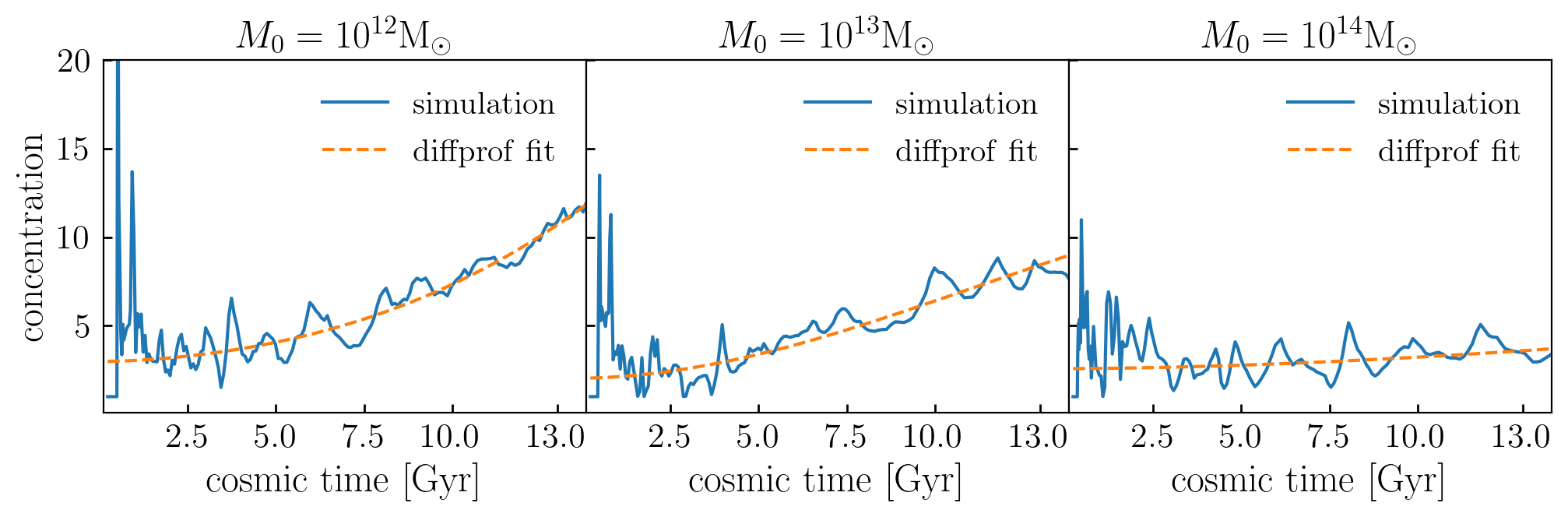}
\caption{{\bf Example fits to concentration histories of individual halos.} The blue curve in each panel shows the concentration history of a dark matter halo taken directly from the merger trees of the Bolshoi-P simulation; the orange curve shows the approximate history based on the \dprof model defined in Eq.~\ref{eqn:ind_model}. Example fits to halos of different masses are shown in each panel.}
\label{fig:individual_conc_fit}
\end{figure*}

All results in the paper pertain to the concentration histories of present-day host halos, defined by Rockstar to be halos at the final snapshot with a {\tt upid} column equal to -1. The boundary of dark matter halos in this catalog was defined according to the virial radius definition, $\rvir$ \citep{bryan_norman98}, and for the value of halo mass we will use $\mpeak,$ the peak historical mass of the main progenitor branch of the halo, as this choice will facilitate our efforts in future work to unify or results with the \dmah model (see \S\ref{sec:discussion} for further discussion). In particular, we will characterize how the concentration histories of halo populations in terms of $\mzero,$ the peak historical halo mass evaluated at redshift zero. For some of the results in the paper, we calculate cross-correlation functions between simulated halos and the density field, using random downsampling of particles from the appropriate snapshot. For Bolshoi-P we use a random downsampling of dark matter particles provided by {\tt halotools} \citep{hearin_etal17_halotools}. Throughout the paper, including the present section, values of mass and distance are quoted assuming $h=1.$ For example, when writing $\mzero=10^{12}\msun,$ we suppress the $\msun/h$ notation and write the units as $\msun.$

\section{Concentration Histories of Individual Halos}
\label{sec:individual_conc}

\begin{figure*}
    \centering
    \includegraphics[width=17cm]{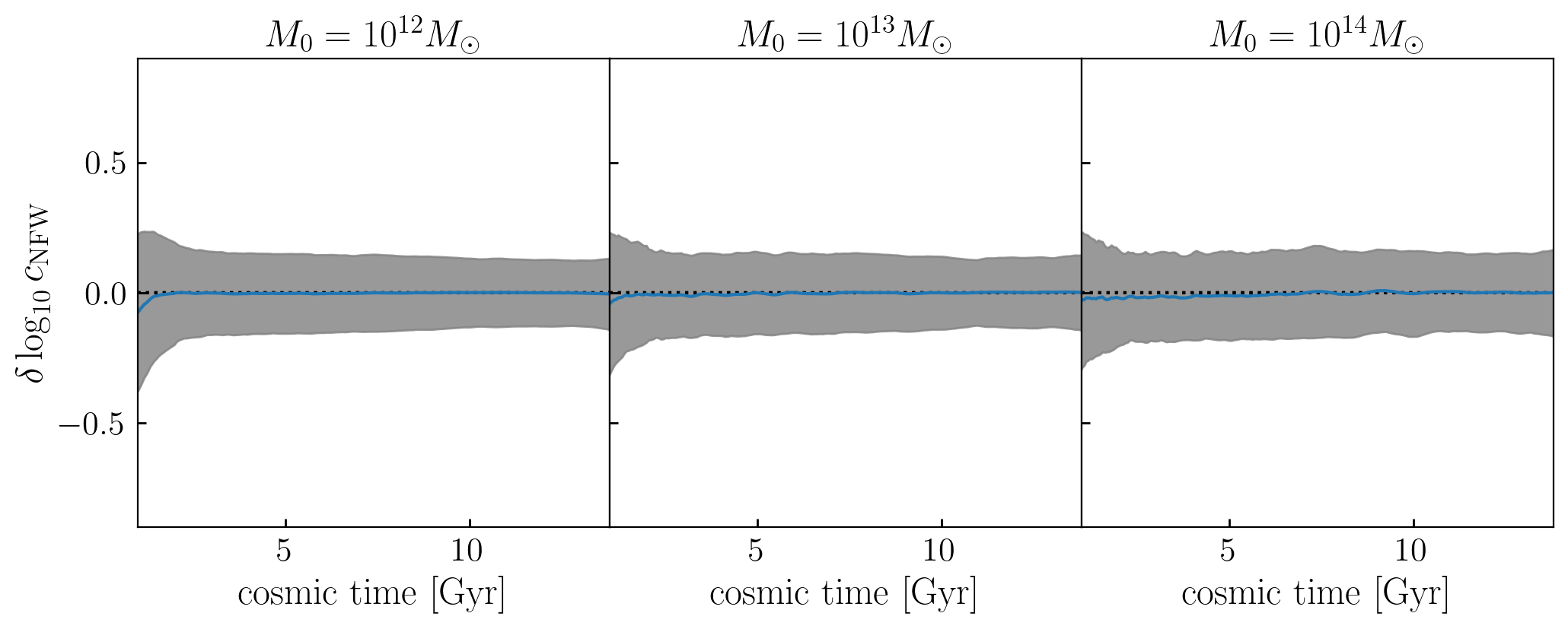}
    \caption{{\bf Residuals of fits to individual halos.} Using a large collection of fits to the concentration histories of simulated halos, the y-axis shows the logarithmic difference between the simulated and best-fit values of $c(t)$; results for halos of different mass are shown in different panels according to the indicated title. The blue curve shows the average residual difference, and the shaded band shows the $1\sigma$ scatter. The figure demonstrates that the \dprof model defined by Eq.~\ref{eqn:ind_model} gives an unbiased approximation to halo concentration history for $t\gtrsim2$~Gyr, with a typical error of $\sim0.1$ dex.
    }
    \label{fig:individual_residuals}
\end{figure*}

The concentration of a dark matter halo is defined in terms of $\rho_{\text{NFW}},$ the NFW model of the radial density profile,
\beq
    \rho_{\rm NFW}(r)\equiv \frac{\rho_s}{(r/r_{\rm s})(1+r/r_{\rm s})^2},
\eeq
where $r$ is the physical distance from the halo center, $r_{\rm s}$ is the scale radius, and $\rho_{\rm s}$ is the normalization \citep[][]{Navarro_1997}. The concentration of the halo is defined as $c \equiv R_{\rm halo}/r_{\rm s},$ where $R_{\rm halo}$ is the halo boundary; as described in \S\ref{sec:sims}, the simulated halo catalogs used throughout this paper are defined with $R_{\rm halo}=\rvir.$

We model the evolution of NFW concentration along the main progenitor branch of an individual halo with a power-law function of cosmic time with a time-dependent index,
\beq
\label{eqn:rolling_plaw}
    c(t) = \cmin10^{\beta(t)}.
\eeq
In Equation \ref{eqn:rolling_plaw}, the quantity $\cmin$ is the asymptotic early-time floor of halo concentration; we model the behavior of $\beta(t)$ using a sigmoid function, $\mathcal{S}(x);$ we use sigmoid functions repeatedly throughout the paper, and so a general definition appears below:
\beq
\label{eqn:sigmoid}
    \mathcal{S}(x) = y_{\rm min} + \frac{y_{\rm max}-y_{\rm min}}{1+\exp[-k(x-x_0)]}.
\eeq
In modeling $\beta(t)$ with a sigmoid, our independent variable $x=\lgg t,$ so that the concentration of an individual dark matter halo evolves according to the following equation:
\beq
\label{eqn:ind_model}
    \lgg c(t) = \lgg\cmin + \frac{\cbl-\lgg\cmin}{1+\exp[-k(\lgg t-\lgg\tauc)]}.
\eeq
In Equation \ref{eqn:ind_model}, the variable $\cbl$ controls the asymptotic behavior at late times; $\tauc$ is the transition time from early- to late-time behavior; and $k$ regulates the rate of the transition between the two regimes. For all results in the paper, we have held $k$ fixed to a constant value of 4; furthermore, we only explore regions of parameter space for which $\cbl>\lgg\cmin,$ so that $c(t)$ can only increase with time. Hereafter, we will refer to this constrained functional form as the \dprof model for the evolution of the NFW concentration of individual halos.

Figure \ref{fig:individual_conc_fit} shows three examples of the concentration histories of dark matter halos, each with a present-day peak mass as indicated by the panel's title. The solid blue curves show the concentration history of the main progenitor halo as taken directly from the simulated merger tree, and the dashed orange curves show the approximation of the fitting function defined in Eq.~\ref{eqn:ind_model}. We give a detailed description of how we find the best-fitting parameters to each halo in Appendix \ref{appendix:individual_conc}.

The concentration histories displayed in Figure \ref{fig:individual_conc_fit} are typical in several respects. At early times, halo concentration fluctuates rather wildly about a constant value of $c\approx3-4,$ and eventually begins to steadily increase towards a present-day value of $c\approx5-15,$ with lower-mass halos tending to have larger present-day values \citep[see, e.g.,][for early work noting these characteristic evolutionary trends]{zhao_etal03}. Although the evolution of the simulated halo becomes smoother as cosmic time progresses, fluctuations in the broad evolution remain present throughout the history of the halos. While the large, early-time fluctuations are at least in part a marker of numerical resolution effects, it is by now well-established that significant excursions from the smooth evolution are caused by minor mergers, and so should be expected even in well-resolved halos \citep[for a recent analysis of this phenomenon, see][]{wang_etal20}. Our differentiable approximation to concentration growth faithfully reproduces the smooth component of halo evolution, but misses the transient features associated with mergers. The remainder of the results in this section illustrates the fidelity with which our model is able to capture the concentration histories of the large statistical samples of the simulated halos described in \S\ref{sec:sims}.

Using the optimization techniques detailed in Appendix~\ref{appendix:individual_conc}, we have identified a set of best-fitting parameters for every halo in the samples described in \S\ref{sec:sims}. On the vertical axis of Figure~\ref{fig:individual_residuals}, we show the logarithmic difference between the simulated and best-fitting concentration history of halos, with results for halos of different masses being shown in different panels. The solid blue curve in each panel shows the average residual difference, and the shaded band shows the $1\sigma$ scatter. For most of cosmic time, the \dprof model gives an unbiased fit to the concentration history of Bolshoi-P and MDPL2 halos, with a typical scatter of $\sim0.1$ dex. The model presents the same level of accuracy and precision for the full range of halo masses we consider, $10^{11.5}\msun\leq\mpeak\leq10^{15}\msun.$ 

\begin{figure}
    \centering
    \includegraphics[width=8cm]{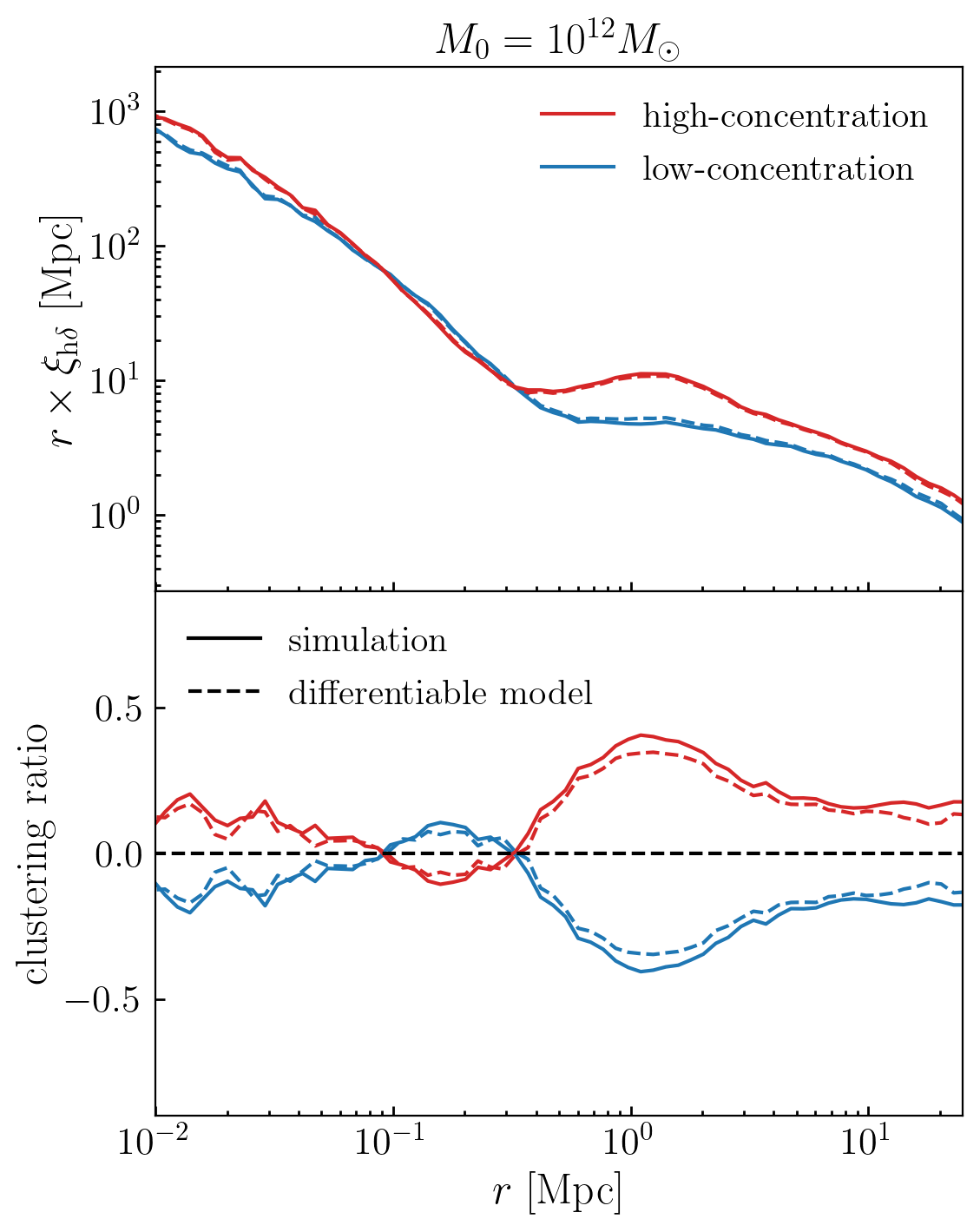}
    \caption{{\bf Concentration-dependence of halo clustering.} In this figure, we focus on a sample of host halos in the Bolshoi-P simulation with $\mpeak=10^{12}\msun,$ a mass range where the well-known phenomenon of {\it halo assembly bias} is particularly strong. We divide the sample in half according to the median concentration value for this mass, and for each subsample we compute $\xi_{\rm h\delta}(r),$ the cross-correlation between halos and dark matter particles. In each panel, solid curves show results for the case where concentration is taken directly from each halo’s simulated merger tree; red curves show $\xi_{\rm h\delta}(r)$ for high-concentration halos, and blue curves show results for low-concentration halos. In the {\it top panel}, we directly plot $\xi_{\rm h\delta}(r),$ and in the {\it bottom panel} we show the fractional difference between the red (blue) curve vs. $\xi_{\rm h\delta}(r)$ for all halos in the sample. Dashed curves show results for the case where concentration is defined by the \dprof approximation. The figure demonstrates that the correlation between halo concentration and the density field is retained when simulated concentration histories are approximated with \dprof.}
    \label{fig:assembly_bias}
\end{figure}

Figure~\ref{fig:individual_residuals} shows that the concentration histories of Bolshoi-P and MDPL2 halos are well approximated by our model, albeit with considerable scatter about the smooth evolution due to transient fluctuations associated with minor mergers. In general, the large-scale clustering of halos exhibits a dependence upon NFW concentration at fixed mass, a phenomenon referred to as {\it secondary halo bias} \citep{gao_etal04}\footnote{Note that ``halo assembly bias'' is an alternative term for this phenomenon, although in the recent literature this has come to specifically refer to the case where the secondary halo property quantifies mass assembly history. In this paper, we will use the general term ``secondary bias" to refer to the specific case of concentration as the secondary parameter; even though this term encompasses the more general case of any arbitrary secondary property, there should be no cause for confusion since in this work we are principally concerned with halo concentration. See, e.g., \citet{salcedo_etal18,mao_etal18}, for further discussion.}. Since the incidence of mergers is correlated with both concentration and with the large-scale density field, a natural question that arises is the extent to which our model can retain the correlation between halo concentration and halo clustering.

\begin{figure*}
    \centering
    \includegraphics[width=12cm]{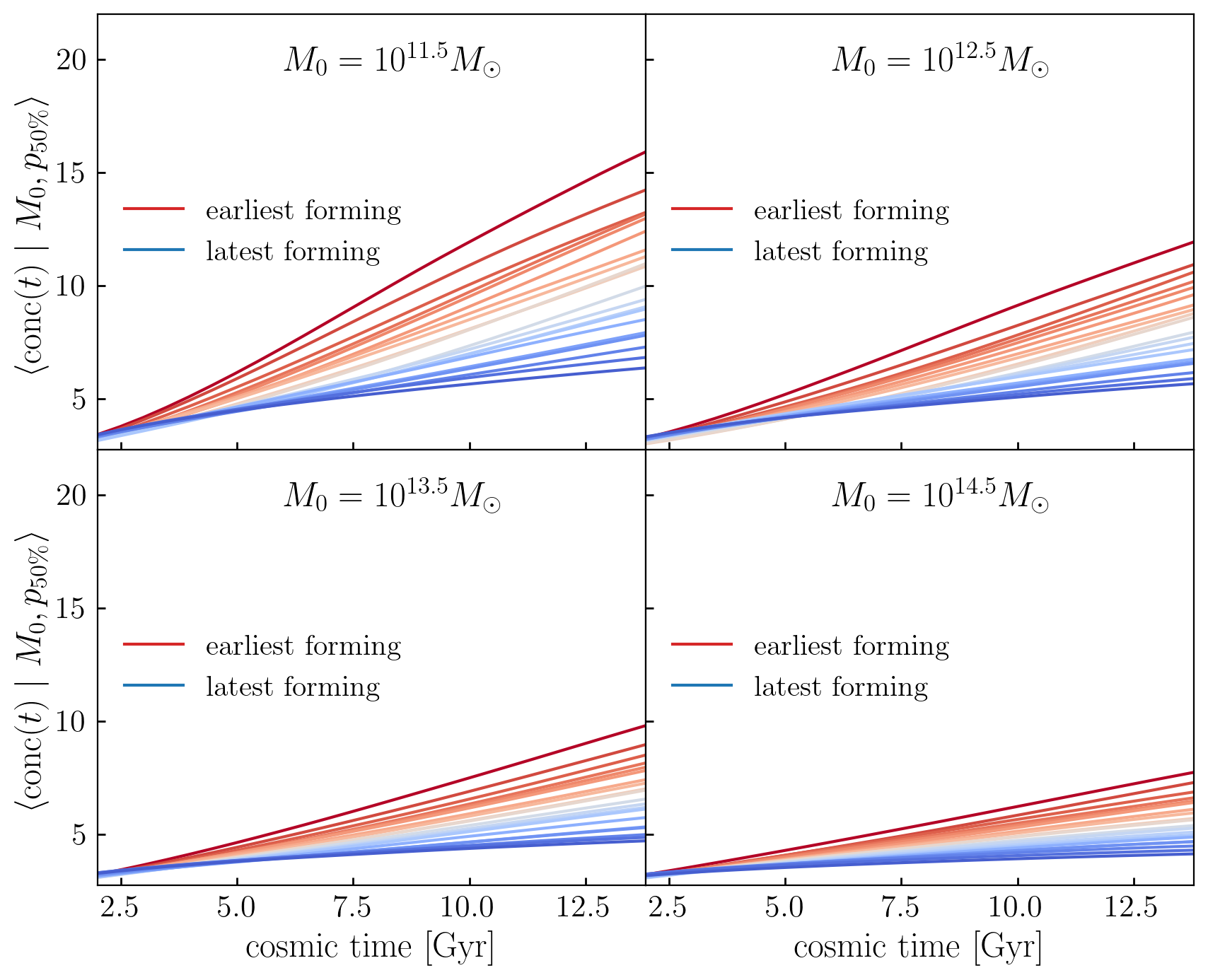}
    \caption{{\bf Average concentration history across time.} Each curve in the figure shows the average concentration history of a sample of halos. Results for halos of different present-day peak masses are shown in different panels. Different colored curves show results for halos with different $\pc,$ which quantifies the {\it percentile} of halo formation time, $\tperc{50},$ conditioned upon $\mzero;$ the reddest curves show results for $\pc=0.1,$ the bluest curves show results for $\pc=0.9,$ and halos with intermediate formation times for their mass are shown with the lighter curves in between the red and blue. For halos of all mass, $c\approx3-4$ at early times, with concentration increasing at later times. Relative to massive halos, the concentration of lower-mass halos is larger and exhibits a tighter connection to mass assembly history.}
    \label{fig:mean_conc_trends}
\end{figure*}

We address this question in Figure~\ref{fig:assembly_bias}. We begin by selecting Bolshoi-P halos at $z=0$ in a narrow mass bin of $\mpeak=10^{12}\msun,$ as the phenomenon of concentration-based secondary bias is particularly strong for halos in this mass range. For the halos in this sample, we calculate $\xi_{\rm h\delta}(r),$ the two-point cross-correlation between halos and dark matter particles at $z=0.$ In each panel of Figure~\ref{fig:assembly_bias}, the red (blue) curve shows results for halos with above-average (below-average) concentration for their mass. On the vertical axis in the top panel, we directly plot $r\xi_{\rm h\delta}(r)$ for the two samples of halos; on the vertical axis in the bottom panel, we plot the fractional difference between the clustering of large- and small-concentration halos relative to the clustering of {\it all} halos in the mass bin; thus the {\it separation} between the red and blue curves in the top panel is quantified by the vertical axis of the bottom panel. Solid curves show the case where the halo sample has been split in half according to the value of concentration at $z=0$ taken directly from the simulated halo catalog; dashed curves show calculations where the halos are split according to the \dprof fit.

On small scales, in the ``1-halo term'', where $r\lesssim\rvir\approx0.2$ Mpc, the difference between the red and blue curves in Figure~\ref{fig:assembly_bias} is a reflection of the dependence of the NFW profile upon concentration. On very small scales, high-concentration halos have larger density relative to low-concentration halos of the same mass, and so the red curve lies above the blue; this difference reverses on spatial scales in the 1-halo term that are larger than the scale radius, $r_{\rm s}\lesssim r\lesssim\rvir,$ where low-concentration halos are denser than high-concentration halos. Finally, the differences between red and blue curves in the 2-halo term on large scales reflect the phenomenon of secondary bias; for halos of this mass, large-scale clustering exhibits a positive correlation with concentration \citep{wechsler_etal2006,mansfield_kravtsov_2020}, and presents a scale-dependent signature at $r\approx1-2$ Mpc \citep{sunayama_etal16}. The dashed curves in Figure~\ref{fig:assembly_bias} are within $\sim5\%$ of the solid curves across the full range of spatial scales, indicating that residual errors in the \dprof approximation to $c(t)$ are largely uncorrelated with the large-scale density field. In Appendix~\ref{appendix:individual_conc}, we demonstrate that most of this difference is driven by extreme outliers in the concentration-mass relation that are likely to be ``splashback halos".


\section{Concentration Histories of Halo Populations}
\label{sec:conc_populations}

In the previous section, we presented a model for the evolution of the NFW concentration of individual dark matter halos. In our model, the evolutionary history of the concentration of a halo is described by three parameters, $\cmin, \cbl,$ and $\tauc,$ with behavior defined by Eq.~\ref{eqn:ind_model}. In this section, we present a model for the probability distribution of $c(t)$ for cosmological populations of halos. Our goal in this section is to develop a model that captures $P(c(t)\vert M_0, \tform),$ the PDF of concentration history across time, and its joint dependence upon $M_0$ and halo assembly time, $\tform.$ Having shown in \S\ref{sec:individual_conc} that the concentration history of individual halos in Bolshoi-P and MDPL2 can be accurately approximated by our parametric fitting function, the approach we take here is to construct a model for the statistical distribution of $\cmin, \cbl,$ and $\tauc.$

In \S\ref{subsec:pop_trends}, we will motivate the functional forms of our model by examining a few basic scaling relations between $c(t),$ $M_0$ and $\tform,$ and in \S\ref{subsec:pop_model_results} we will present our model and assess the accuracy with which it can capture both the {\it average} concentration history of halo populations, as well as the {\it diversity} of evolutionary histories. In the main body of the paper, we will focus primarily on the formulation of our model and the principal demonstrations of its accuracy; a full account of our optimization procedure and attendant details can be found in the appendices together with our publicly available code.

\subsection{Basic trends of halo populations}
\label{subsec:pop_trends}

In this section, we motivate the formulation of our model for the concentration histories of halo populations. We will characterize the $\tform$-dependence of concentration history in terms of $\tform\equiv\tperc{50},$ the half-mass time at which the main progenitor mass of the halo first exceeds $M_0/2.$ Because the average value of $\tperc{50}$ itself depends upon $M_0,$ we find it useful to quantify halo assembly history in terms of $\pc\equiv P(<\tperc{50}\vert M_0),$ the mass-conditioned cumulative distribution of $\tperc{50}.$ Thus by definition, we have $0<\pc<1,$ with smaller values of $\pc$ corresponding to halos with early formation times for their mass.

\begin{figure}
    \centering
    \includegraphics[width=8cm]{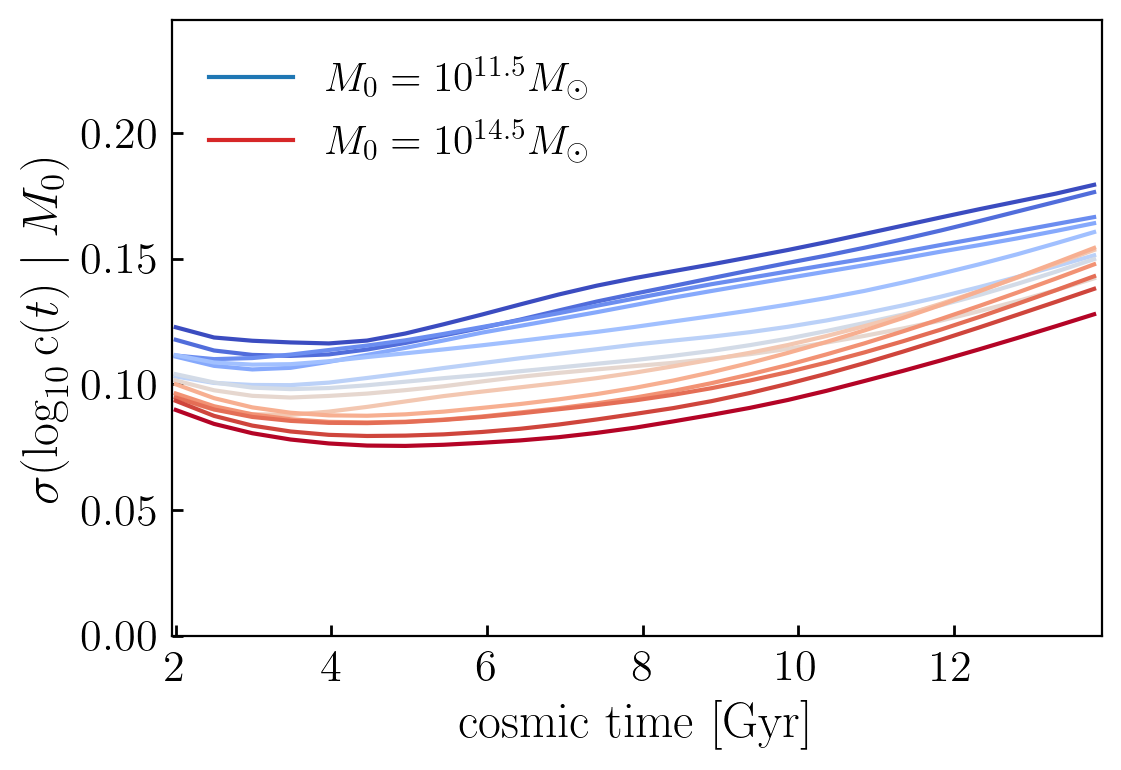}
    \caption{{\bf Scatter in concentration history across time.} Each curve in the figure shows the level of scatter in halo concentration across time. Different colored curves show results for halos of different present-day peak masses, using a color gradient that varies logarithmically from the low-mass end in blue to the high-mass end in red. For halos of all mass, scatter in concentration increases with time. At all cosmic times, lower-mass halos have a larger scatter in concentration relative to massive halos.}
    \label{fig:std_conc_trends}
\end{figure}

In Figure~\ref{fig:mean_conc_trends}, we show how the average history of concentration depends upon $M_0$ and $\pc.$ Each panel shows results for a sample of halos of different mass, as indicated by the in-panel annotation. Red curves in each panel show results for $\pc\approx0,$ and blue curves show results for $\pc\approx1.$ In this figure and throughout the paper, we use the Bolshoi-P simulation for halos with $M_0<10^{13.5}\msun,$ and the MDPL2 simulation for results based on halos at higher mass.

Figure~\ref{fig:mean_conc_trends} illustrates many of the key features of the statistical distribution of concentration histories that we wish to capture with our model for halo populations. For halos of all mass and assembly history, at high redshift we see that $c(t)\approx3-4$, and that $c(t)$ tends to increase over time. At late times, lower-mass halos have higher concentrations relative to higher-mass halos. For halos of all mass, earlier-forming halos have higher concentrations than later-forming ones. By comparing the top-left to the bottom-right panels, we can also see that the concentrations of lower-mass halos present a much stronger dependence upon halo assembly history relative to massive halos; this is sensible, since lower-mass halos are highly susceptible to environmental influence \citep[e.g.,][]{mansfield_kravtsov_2020}, whereas cluster-mass halos dominate the tidal field of their environment.

\begin{figure}
    \centering
    \includegraphics[width=8cm]{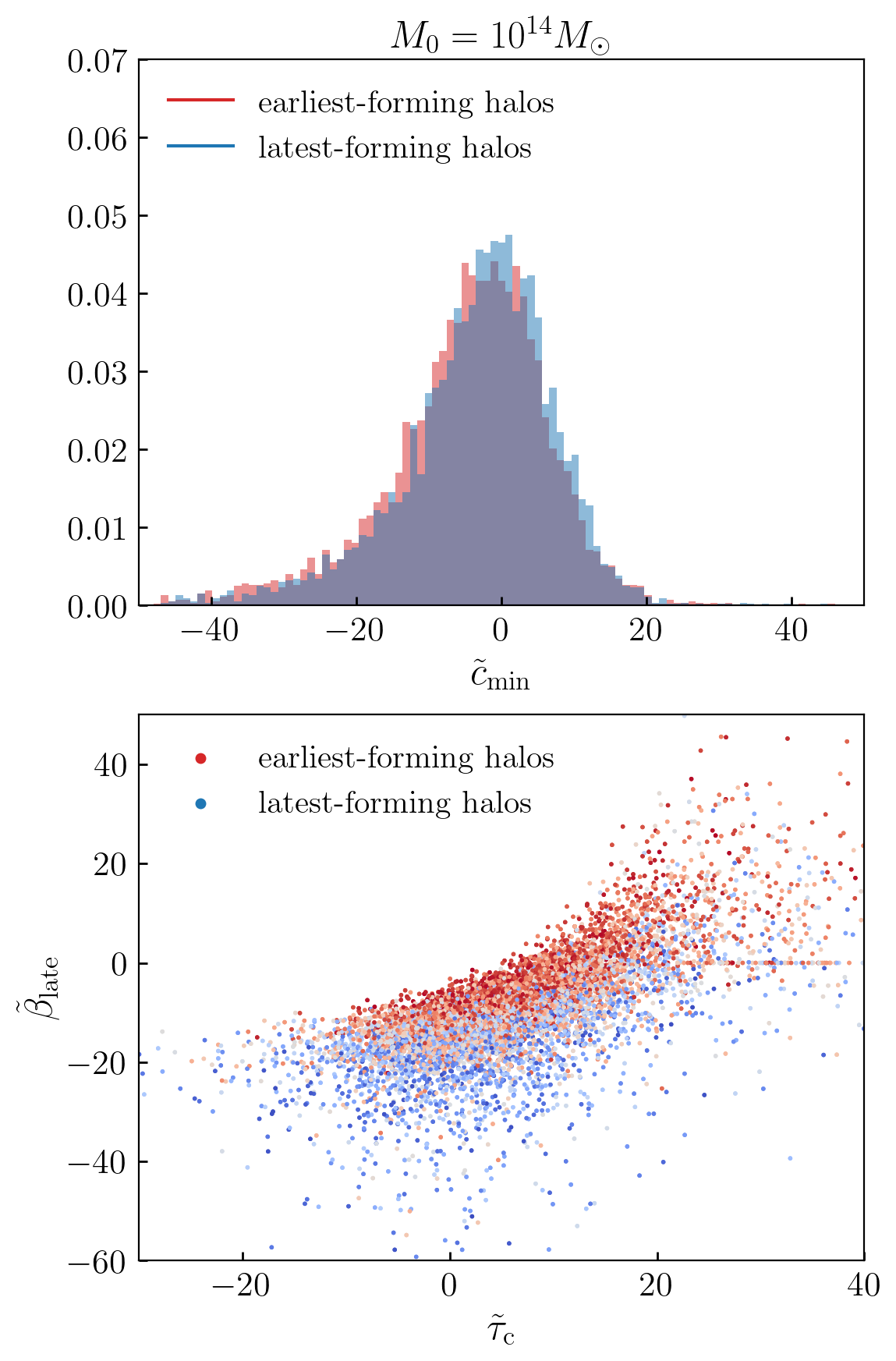}
    \caption{{\bf Distribution of parameters for concentration evolution.} The figure shows the distribution of best-fitting parameters of our model for the evolution of concentration of individual halos, $\cmin, \cbl,$ and $\tauc$ (see Eq.~\ref{eqn:ind_model}), focusing on a sample of host halos in the MDPL2 simulation with $M_0\approx10^{14}\msun.$ As described in the text, for each parameter, $x,$ the figure shows $\tilde{x},$ which has a nonlinear but monotonic relationship to the parameter. In the top panel, we show a histogram of $\cminprime,$ which has an approximately Gaussian shape with minimal dependence upon assembly time. The bottom panel illustrates the correlated relationship between $\cblprime-\taucprime,$ which exhibits a strong correlation with halo assembly. The distributions shown here motivate the model for the concentrations of halo populations presented in \S\ref{subsec:pop_model_results}.}
    \label{fig:param_distributions}
\end{figure}

In Figure~\ref{fig:std_conc_trends}, we show how the scatter in halo concentration evolves across time, with results for halos of different mass shown with different colors as indicated in the legend. The figure shows that for halos of all mass, scatter in concentration tends to increase with time. This is consistent with the results displayed in Figure~\ref{fig:mean_conc_trends}, as well as with the physical picture of concentration evolution reviewed in \S\ref{sec:intro}. At early times, halo assembly is firmly in the fast-accretion regime, and concentration takes on a nearly constant value of $c\approx3-4;$ as cosmic time progresses, halo growth slows down, and concentration tends to increase in a manner that is subject to significant environmental influence, leading to larger variance at later times. This environment-correlated variance is more pronounced in lower-mass halos, which is reflected by the bluer curves lying above the redder curves in Figure~\ref{fig:std_conc_trends}.

Figures~\ref{fig:mean_conc_trends}-\ref{fig:std_conc_trends} display the basic trends that we wish to capture with our model for  the concentration histories of halo populations, quantified by $P(c(t)\vert\mzero,\pc).$ We will formulate our model in terms of $P(\cmin, \cbl, \tauc\vert\mzero,\pc),$ the statistical distribution of the best-fitting parameters appearing in Eq.~\ref{eqn:ind_model} that describe the concentration histories {\it individual} halos.

\begin{figure*}
    \centering
    \includegraphics[width=15cm]{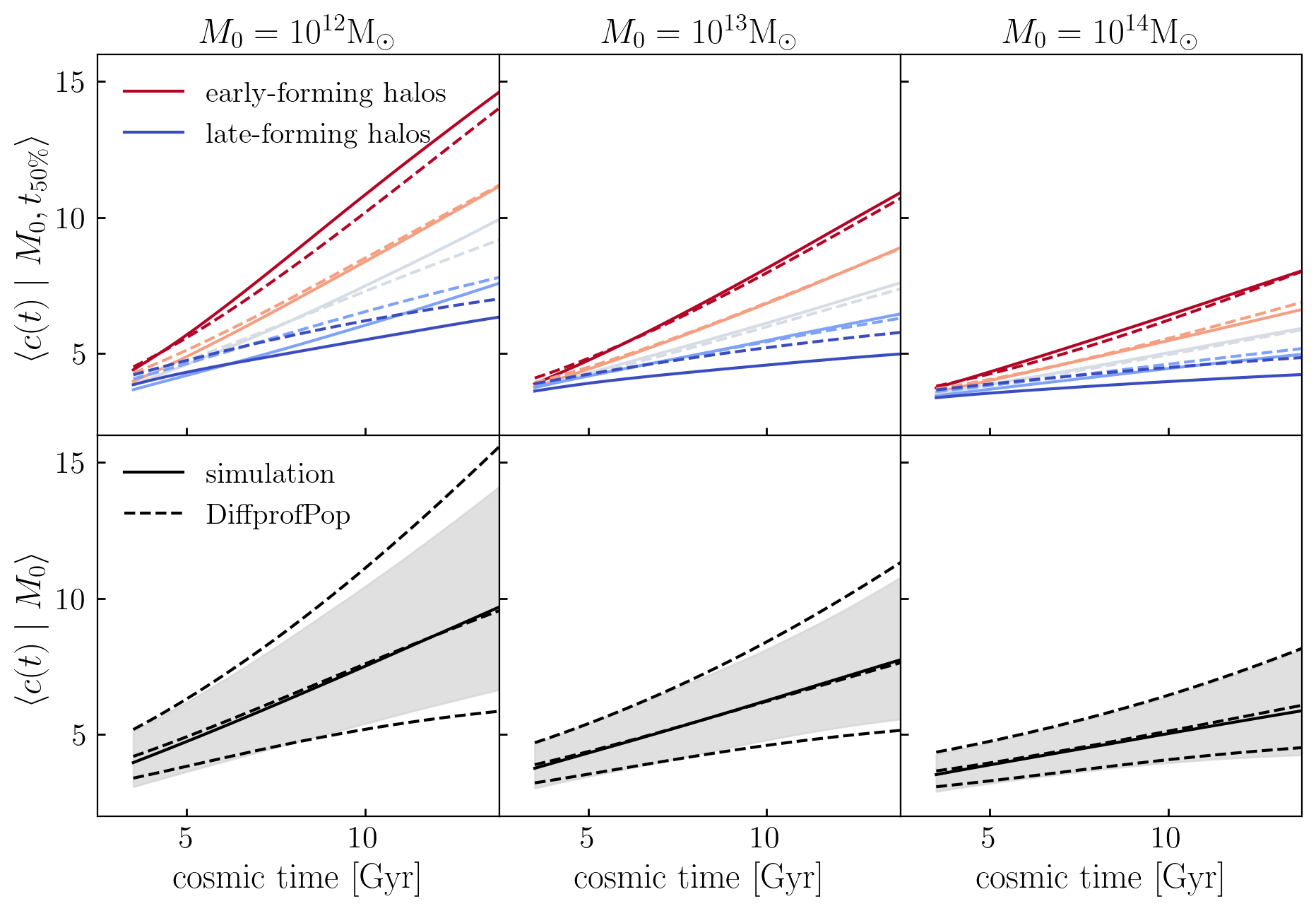}
    \caption{{\bf Concentration histories of halo populations.} The figure shows $P(c(t)\vert\mzero,\pc),$ the statistical distribution of $c(t)$ for populations of halos, shown as a function of halo mass, $\mzero,$ and the {\it percentile} of halo formation time, $\pc\equiv P(<\tperc{50}\vert\mzero).$ Each of the three columns of panels shows results for halos of different present-day peak mass, $\mzero.$ In each panel, solid curves show concentration histories of simulated halos, and dashed curves show the corresponding predictions of our best-fitting model. In the top row of panels, we show the evolution of the average concentration of halos as a function of mass and assembly time; the reddest curves show results for $\pc=0,$ the bluest curves show results for $\pc=1,$ and halos with median formation times are shown with the lighter curves in between. In the bottom row of panels,  we average over the $\tperc{50}$-dependence and show how the mean concentration evolves with time, and additionally show the scatter in concentration at fixed mass, $\sigma(c(t)\vert\mzero).$ The figure shows that our model for halo populations can capture both the {\it average} evolution of concentration, as well as the {\it diversity} of concentration trajectories across time, including dependence upon both halo mass and assembly history.}
    \label{fig:conc_pdf_model_results}
\end{figure*}

In Figure~\ref{fig:param_distributions}, we motivate the formulation for our model of $P(\cmin, \cbl, \tauc\vert\mzero,\pc)$  by illustrating two different cross-sections of the statistical distribution of $\cmin, \cbl,$ and $\tauc,$ focusing on a sample of cluster-mass halos in the MDPL2 simulation. As described in Appendix~\ref{appendix:individual_conc}, when fitting individual concentration histories with our model, we first apply a nonlinear transformation  to each of the variables appearing in Eq.~\ref{eqn:ind_model}, as this enforces physical constraints in the approximation to each halo's concentration history. Each of the transformations we use is monotonic, so that larger values of each variable, $x,$ correspond to larger values of $\tilde{x}.$ The variables $\cminprime, \cblprime,$ and $\taucprime$ are the actual quantities that we programmatically vary when seeking best-fitting approximations to the concentration histories of individual halos, and these are the variables appearing on the axes in Figure~\ref{fig:param_distributions}.

In the top panel of Figure~\ref{fig:param_distributions}, we show two histograms that display the behavior of $P(\cminprime\vert M_0, \pc).$  The red (blue) histogram shows results for halos with formation times in the bottom (top) quartile of $\pc.$ From the top panel, we can see that $P(\cminprime\vert M_0, \pc)$ has an approximately Gaussian shape that is essentially independent of $\pc.$ In the bottom panel of Figure~\ref{fig:param_distributions}, we illustrate the two-dimensional distribution of $P(\cblprime, \taucprime\vert M_0, \pc),$ where we have color-coded the scattered points according to $\pc.$ Referring back to Equation~\ref{eqn:ind_model}, larger values of $\cbl$ correspond to halos with higher concentrations today, and so in the bottom panel we can see a manifestation of the result that at fixed halo mass, earlier-forming halos tend to have higher concentrations relative to later-forming halos of the same mass. In the next section, we will build a model for the statistical distributions shown in Figure~\ref{fig:param_distributions}, and show that our model is able to capture the key evolutionary trends illustrated in Figures~\ref{fig:mean_conc_trends}-\ref{fig:std_conc_trends}.

\subsection{Halo population model}
\label{subsec:pop_model_results}

In this section, we present our model for the distribution of concentrations of populations of dark matter halos, $P(c(t)\vert\mzero,\pc).$ As described in \S\ref{subsec:pop_trends}, we approach this problem through a parametrized description of the probability distribution of $\cmin, \cbl,$ and $\tauc.$ Thus, the fundamental unit of our model is a smooth trajectory in halo concentration defined by Eq.~\ref{eqn:ind_model}, and our model characterizes the PDF of these trajectories, including correlations with halo mass and assembly.

As mentioned in \S\ref{subsec:pop_trends} and discussed in detail in Appendix~\ref{appendix:individual_conc}, we will define our model in terms of $\cminprime, \cblprime,$ and $\taucprime,$ each of which is a transformed variable of the quantities appearing in Eq.~\ref{eqn:ind_model}. The purpose of these transformations is to enforce physical constraints on the fits to individual concentration histories, in particular that the \dprof smooth approximations of $c(t)$ are bounded and non-decreasing. 
We find little to no correlation between $\cminprime$ and either $\mzero$ or $\pc,$ and so we model $\cminprime$ to be distributed as a Gaussian with uncorrelated scatter. We model $\cblprime-\taucprime$ using a two-dimensional Gaussian with a mean and covariance matrix that jointly depends upon $\mzero$ and $\pc.$ We give a full description of our functional forms in Appendix~\ref{appendix:diffprofpop_definition}.

Figure~\ref{fig:conc_pdf_model_results} displays our model's ability to capture the distribution of concentration histories, $P(c(t)\vert\mzero,\pc).$ The top row of panels shows $\langle c(t)\vert\mzero,\pc\rangle,$ the evolution of the average concentration of halos as a function of mass and the rank-order percentile of $\tperc{50}.$ Each of the three columns of panels shows results for halos of different mass. Within each panel on the top row, there are four colored curves, each showing the average histories of halos with different values of $\pc=(0.2, 0.4, 0.6, 0.8);$ redder (bluer) colors pertain to halos with earlier (later) assembly times for their mass. Solid curves show the concentration histories of simulated halos, and dashed curves show  predictions for $P(c(t)\vert\mzero,\pc)$ deriving from our parametrized approximation. Each curve showing simulated halo histories was computed using tens of thousands of halos. The \dprofpop model describes the evolutionary trends of $c(t)$ reasonably well for halos of all mass that we consider, including the $\pc$-dependence. Our model is comparatively less accurate for very early-forming halos of lower mass, although it is unclear to what extent this shortcoming is due to splashback subhalos that should be excluded from the simulation data; we refer the reader to \S\ref{subsub:outlier_halos} for further discussion of this issue.

In the bottom row of panels of Figure~\ref{fig:conc_pdf_model_results}, we plot $\langle c(t)\vert\mzero\rangle,$ and additionally show the scatter in concentration at fixed mass, $\sigma(c(t)\vert\mzero).$ For simulated halos, the $\mzero$-averaged concentration is shown with the solid black curve, and the gray band shows the $1\sigma$ scatter; the corresponding predictions of our model are shown with dashed curves. Taken together, the panels of Figure~\ref{fig:conc_pdf_model_results} show that our model for the concentration of halo populations can capture both the {\it average} evolution of concentration, as well as the {\it diversity} of concentration trajectories across time, including dependence upon both halo mass and assembly history.


\section{Discussion}
\label{sec:discussion}

We have presented a new model for the evolutionary history of dark matter halo concentration, $c(t).$ In our model,
the concentration of an individual dark matter halo evolves as a power-law function of cosmic time with a time-dependent index, $c(t)=c_{\rm min}10^{\beta(t)},$ where $\beta(t)$ is a sigmoid function that transitions halo concentration growth from constant, early-time behavior, to smoothly increasing late-time behavior (see Equations~\ref{eqn:rolling_plaw}-\ref{eqn:ind_model}). The results in \S\ref{sec:individual_conc} demonstrate that for host halos identified with Rockstar in the Bolshoi-P and MDPL2 simulations over a wide range of present-day peak mass, $10^{11.5}\msun\lesssim\mzero\lesssim10^{15}\msun,$ our model accurately approximates the concentration of individual halos for $t\gtrsim2$~Gyr. We have additionally presented a new model for the concentration history of halo populations; our model captures both {\it average} trends between concentration and halo mass and assembly, as well as the {\it diversity} of concentration histories of simulated halos.

In \S\ref{subsec:diffpop}, we describe the broader context of our modeling approach, discussing how \dprof is a specific example of a framework for differentiable modeling populations of galaxies and halos, and how this framework relates to current approaches in the literature. We discuss the limitations of our model in \S\ref{subsec:limitations}, and future extensions in \S\ref{subsec:future}.

\subsection{Differentiable Population Modeling Approach}
\label{subsec:diffpop}

As described in detail in the appendices, our characterization of the diversity in halo concentration is based on the differentiable population modeling technique described in \citet{Hearin_2021}. When using this framework to model some galaxy or halo property, $X,$ the core modeling ingredient is a family of fitting functions, $\mathcal{F}_{\theta_X}(t),$ parametrized by $\theta_{X},$ which approximates the time-evolution $X(t)$ of an {\it individual} object. For the cosmological distribution of $X,$ a separate model parametrized by $\Psi_X$ characterizes $P_{\Psi_X}(\theta_{X}),$ the statistical distribution of $\theta_{X},$ which together with $\mathcal{F}_{\theta_X}(t)$ provides a model for the probability distribution of $X$ across time. For models formulated in this fashion, predictions for $X(t)$ of individual objects are deterministically specified by $\theta_{X},$ and predictions for populations of $X(t)$ are deterministically specified by $\Psi_{X};$ thus by implementing these two modeling ingredients in a library for automatic differentiation such as JAX \citep{jax2018github}, exact gradient information becomes available when optimizing the parameters. 

In \citet{Hearin_2021}, this differentiable framework was used to model the evolution of the mass of individual dark matter halos; the quantity $X(t)=\mpeak(t)$ was approximated with the \dmah functional form, $\mathcal{F}_{\theta_{\rm MAH}}(t),$ whose behavior is specified by three parameters: $\theta_{\rm MAH}=\{\aearly, \alate, \tauh\};$ the distribution of halo mass across time is described by \dmahpop, a family of fitting functions parametrized by $\Psi_{\rm MAH}$ that approximates the statistical distribution of $P(\aearly, \alate, \tauh\vert\mzero).$ The parameters $\Psi_{\rm MAH}$ were calibrated using merger trees in high-resolution cosmological N-body simulations, so that \dmahpop accurately approximates $P(\mpeak(t)\vert\mzero).$

In \citet{diffstar_alarcon_etal23}, the same framework was used to model the star formation history (SFH) of individual galaxies. The quantity $X(t)=\dot{M}_{\star}(t)$ was approximated with the \dstar functional form, whose behavior is specified by eight parameters, $\theta_{\rm SFH};$ the \dstar model was shown to accurately approximate the SFHs of individual galaxies in the UniverseMachine \citep{universemachine_behroozi19} and TNG \citep{tng_pillepich_2018} simulations. In ongoing work, we are building \dstarpop: a parametrized model for $P(\theta_{\rm SFH}\vert\theta_{\rm MAH}),$ which will enable differentiable predictions for the statistical distribution of SFH across time for cosmological samples of galaxies.

In the present work, we study $X(t)=c(t),$ the time-evolution of the NFW concentration of individual dark matter halos; we approximate $c(t)$ using the \dprof model defined in Eq.~\ref{eqn:ind_model}, which is controlled by three parameters: $\thetanfw=\{\cmin, \cbl, \tauc\}.$ We then characterize the cosmological abundance of halo concentration using \dprofpop, a model for the probability distribution $P(\thetanfw)$ that includes dependence upon both halo mass and assembly history:
\beq
\label{eq:probc}
P(c(t)\vert\mzero,\tform)=P(\cmin, \cbl,\tauc\vert\mzero,\tform),
\eeq
where $\mzero$ is the present-day peak halo mass, and $\tform=\tperc{50}.$ 

In typical approaches to modeling halo concentration, diversity in $c$ at fixed $\mhalo$ arises in one of two ways. First, in models that directly parameterize the $c$-$\mhalo$ relation \citep[e.g.,][]{Bhattacharya_etal13,dutton_maccio14}, scatter at fixed mass and redshift is simply considered to be a random variable distributed as a log-normal; this class of models can accurately approximate the distribution of concentration as a function of mass and redshift, but is not equipped to characterize the time-evolution of the concentration of individual halos. There is a second class of models that instead parameterizes the evolution of halo concentration \citep[e.g.,][]{Zhao_2009,ludlow_etal13,correa_etal15}; these models {\it do} have the predictive power to describe the evolution of individual halo concentration (as well as variations in cosmology, see below), and in each of these models, scatter at fixed mass and redshift arises exclusively due to variations in some particular summary statistic of halo assembly history, such as $\tform=\tperc{4},$ or $\tform=t_{-2}.$

Diversity in halo concentration at fixed mass arises in our \dprofpop model through two separate channels. First, there is a great diversity of trajectories in time by which halos attain the same final mass $\mzero.$ In \dprofpop, the statistical distribution $P(\theta_{\rm NFW})$ presents a strong dependence upon halo formation time, $\tform,$ which gives rise to the bulk of the scatter in the $c$-$\mhalo$ relationship. Second, in contrast to the models discussed above, \dprofpop captures a {\it distribution} of halo concentrations at fixed values of halo mass and formation time. 

The \dprof model approximates the {\it smooth} evolutionary history of $c(t),$ but not transient fluctuations associated with the merging of substructure, and so our models do not capture the contribution of such fluctuations to the variance in $c(t)$ at fixed halo mass. Transient merging events have been shown to significantly influence the value of concentration at a particular time \citep{wang_etal20,lucie_smith_etal22}, and smooth approximations to the full assembly history only contain a portion of the available information \citep{mendoza_multicam_2023}; an additional modeling ingredient beyond what we introduce here would be required to capture these fluctuations. Encouragingly, in approximating $c(t)$ of simulated halos with the smoothly-evolving \dprof, the residual errors appear to be largely uncorrelated with the large-scale density field (see Figures~\ref{fig:assembly_bias} \& \ref{fig:assembly_bias_percentiles}). This indicates that even though our model is a simplification of the full merger tree, theoretical predictions for large-scale structure may not suffer from appreciable biases associated with using a smoothly-evolving approximation to $c(t).$ We will explore this and other related topics in the future work outlined in \S\ref{subsec:future}, and in the next section \S\ref{subsec:limitations}, we discuss other shortcomings of our model and caveats to our conclusions.

\subsection{Limitations and Caveats}
\label{subsec:limitations}

\subsubsection{Cosmology dependence}
As our paper is the first attempt at applying the differentiable population modeling technique to halo concentration, we have employed numerous simplifying assumptions that limit the predictive power of \dprof relative to more mature modeling efforts. For example, the two simulations we used were both run with cosmological parameters similar to \citet{planck14b}, whereas more mature modeling efforts have the capability to capture the cosmology-dependence of halo concentration \citep[e.g.,][]{Zhao_2009,correa_etal15,lopez_cano_etal2022}. One approach to incorporating cosmology dependence would be to adapt techniques utilizing physically-motivated rescalings \citep[e.g.,][]{angulo_white_2010,renneby_etal18,arico_angulo_etal20}, or alternatively, to use a large suite of simulations to develop an emulator-type approach \citep[as in, for example,][]{heitmann_etal16_mira_titan,derose_etal19_aemulus1,nishimichi_etal19_dark_quest}. While the emulator approach would have stringent demands for merger trees and high mass-resolution, Gpc-scale high-resolution simulations with merger trees are becoming increasingly common \citep{ishiyama_etal21_uchuu,frontiere_etal22}, as are suites of simulations that include merger trees \citep{contreras_etal20}.

\subsubsection{Baryonic effects}
\label{subsec:baryonic_effects}
A second simplifying assumption used in our analysis is that we have restricted attention to gravity-only N-body simulations, although it is well known that baryonic effects have a significant impact on the internal structure of dark matter halos \citep{gnedin04,kazantzidis04,jing_etal06,rudd_etal08,duffy10}. There has been considerable recent progress in the development of large suites of simulations that span a wide range of baryonic effects \citep{villaescusa_navarro_etal20_camels1}, and efforts along these lines have already uncovered a wealth of information about the relationship between feedback and halo internal structure \citep{chua_etal21,anbajagane22}. As baryonic effects have an important influence upon the cluster mass--observable relation \citep[e.g.,][]{nagai07,battaglia12,lebrun14,truong18}, it would be both interesting and well-motivated to extend the \dprof model to incorporate dependence upon baryonic feedback.

\subsubsection{Formation time definition}

We note that in principle, we could have instead elected to build a model that depends not just upon a particular definition of $\tform,$ but instead with joint dependence upon all three \dmah parameters, $\aearly, \alate,$ and $\tauh.$ Here, we have opted for a simpler approach that essentially uses a single measure of halo formation time as a one-dimensional approximation to the three-dimensional dependence of $c(t)$ upon $\aearly, \alate,$ and $\tauh.$ Since the \dmah model accurately characterizes the probability distribution $P(\mpeak(t)\vert M_0),$ the more complex three-dimensional approach has the potential to capture the dependence of $c(t)$ upon the full distribution of halo assembly trajectories. On the other hand, in the simpler approach taken here, we have used the {\it formation time percentile}, $\pc,$ as the one-dimensional variable approximating halo assembly, which has the potential advantage of being more robust to modifications from cosmology or baryonic feedback, as these and other effects may only alter the numerical value of $\tperc{50},$ without changing dependencies upon rank-order. In future work, we will explore this three-dimensional generalization, as well as other methods for quantifying the mutual covariance between concentration evolution and halo assembly.

\subsubsection{Halo boundary definition}
\label{subsub:outlier_halos}
As pointed out in \S\ref{subsec:diffpop}, the \dprofpop model is comparatively less accurate in making predictions for the earliest-forming halos of mass $\mzero\lesssim10^{12}{\rm M_{\odot}}.$ As shown in Figure~\ref{fig:assembly_bias_percentiles} of Appendix~\ref{appendix:individual_conc}, the residual errors of the \dprof approximation appear to be mildly correlated with the halo--matter cross-correlation function, particularly on scales $r\approx1-2$ Mpc. These two outlier populations of halos are heavily overlapping, since the $c$--$\tform$ connection is fairly tight for halos of this mass. A sizable fraction of such halos are likely to be splashback subhalos that are only temporarily outside the virial radius of their host \citep{mansfield_kravtsov_2020,diemer_flybs_2021}; thus the shortcoming of our model in this regime might be ameliorated by adopting simulation data based on a halo boundary definition that is better physically-motivated than the virial radius \citep[e.g.,][]{diemer_dynamics_based_halodef_2022_paper1,garcia_better_halo_boundary_defn_2022}; otherwise an additional modeling ingredient would need to be introduced to capture this subpopulation. We leave such an investigation as a task for future work based on a higher-resolution suite of simulations.

\subsubsection{Numerical resolution}

When calibrating our model for individual concentration histories in \S\ref{sec:individual_conc}, as well as our model for the concentrations of halo populations in \S\ref{sec:conc_populations}, we have focused on the mass range $10^{11.5}\msun\leq\mpeak\leq10^{15}\msun,$ as this range of masses contains thousands of halos that are resolved by over $2000$ particles in the N-body simulations we use. However, we note that even this restricted mass range pushes the resolution limits of our simulations. For example, Bolshoi-P halos at the low-mass end of this range have only $\sim100$ particles at the high-redshift end of our target cosmological epoch. Broad guidelines from previous work has shown that simulated halos must have at least 200 particles {\it within the scale radius} in order to have a reasonably well-measured concentration \citep{klypin_etal2001}. But for science objectives such as precision cosmology that require strict convergence, a large body of previous work has shown more methodical criteria than this is warranted. Methodical resolution studies typically estimate a convergence radius, $r_{\rm conv},$ that specifies the radius at which the density profile is subject to numerical effects such as artificial relaxation; the value of $r_{\rm conv}$ varies from halo to halo, and defines minimum radius used when fitting each profile \citep[see, e.g.,][]{neto_etal07,ludlow_etal19,brown_etal2022}. 

The basic goals of the present paper are more modest. Here we introduce a novel modeling framework for approximating the evolution of halo concentration, and we restrict our scope to exploring the scientific potential of our approach. As discussed in Appendix \ref{appendix:individual_conc}, we have restricted our halo mass range, as well as the range of cosmic time $t>2$~Gyr, in an effort to have target halo data based on publicly available simulations that {\it i}) has broadly reasonable concentrations, and {\it ii}) spans a sufficiently wide range of mass and time to warrant development of a population model. However, a dedicated convergence study will be required before we are able to transform \dprof into a precision tool. As indicated in \citet{mansfield_avestruz_2021}, such a convergence study would require an analogous effort to what has been done to calibrate models of the halo mass function \citep[e.g.,][]{jenkins_etal01_mass_function,tinker_etal08_mass_function,mcclintock_etal19_mass_function,bocquet_etal20_mass_function}. This effort would be facilitated by the public availability of recent suites of N-body simulations such as Sympony \citep{symphony_nadler_2022} and MORIA \citep{diemer20_sparta3}. We consider our results based on Bolshoi-P and MDPL2 to be a promising proof-of-principle that motivates a dedicated convergence study in future work.


\subsection{Future Work}
\label{subsec:future}

In future work on the population modeling approach outlined in \S\ref{subsec:diffpop}, \dprof will provide a basic ingredient in joint predictions for the density field traced by the galaxies and gas in dark matter halos. In one such application that relies on high-resolution simulations with data products that include merger trees, \dprof is used in a pre-processing step in which $\thetanfw$ provides an approximate description of $c(t)$ for every halo. This class of application makes no use of \dprofpop; instead, for each simulated halo, rather than characterizing $c(t)$ by its tabulation at each of the $\sim200$ snapshots, $c(t)$ is systematically replaced by its 3-dimensional, $\thetanfw$-based approximation. Forward modeling with survey-scale simulations is extremely memory intensive, and so \dprof reduces the memory footprint of these predictions. Our formulation in terms of individual halo trajectories of $c(t)$ may also simplify future modeling of the time-evolution of baryonic effects on halo profiles \citep[e.g.,][]{rudd_etal08,schneider_teyssier_2015}.

In a separate set of applications, \dprofpop is used to generate a synthetic population of halos with a {distribution} of $c(t)$ that is statistically representative of simulated halo histories. Downstream ingredients for observables such as Compton-y are then based on these synthetic histories. In this second set of applications, in which \dprofpop generates synthetic halo populations, downstream predictions are typically only made for one-point functions such as a mass-observable scaling relation and its scatter. This application of \dprofpop is directly analogous to the way some semi-analytic models \citep[e.g.,][]{somerville_primack_1999,benson_galacticus_2012} predict quantities such as the galaxy luminosity function using merger trees generated with either extended Press-Schechter methods \citep{bond_etal91_eps,bower_1991,1acey_cole_1993}, or with machine learning \citep{nguyen_etal_2023_florah}. Recent work introducing the TOPSEM model of galaxy disks and bulges has used the DiffmahPop model in this fashion \citep{boco_etal_2023_topsem}.

In the present work, we have defined the internal structure of dark matter halos in terms of the NFW approximation of a spherically symmetric radial distribution, but in reality, halos are not perfect spheres: their internal structure is more accurately described by a triaxial ellipsoid characterized by an ellipticity and a prolaticity \citep{jing_suto_2002}, both of which vary with halo mass and redshift \citep[e.g.,][]{allgood_etal06,bonamigo_etal15,lau_etal21}. Just as with concentration, there is a significant scatter in halo ellipticity at fixed halo mass, and this scatter is strongly correlated with halo assembly history \citep{despali_etal14,chen_etal20}. In a companion paper to the present work, we will generalize the \dprof model to characterize the {\it joint} evolution of halo concentration and triaxility, again following a differentiable population modeling approach to capture physically realistic covariance with halo mass and assembly history.

Even when neglecting ellipsoidal deviations from spherical symmetry, the NFW density profile fails to capture the well-known ``splashback" feature of the outer halo profile \citep{more_diemer_kravtsov15_splashback,diemer17_sparta,mansfield_etal17_shellfish,oneil_etal21_splashback_hydro}. The splashback radius of a halo is tightly connected to its recent mass accretion history \citep{diemer_kravtsov14_outer_profiles,diemer_etal17_sparta2}, which has motivated numerous efforts to model and measure this signature in observations \citep{umetsu_diemer17,surcher_more_19_splashback_planckSZ,xhakaj_etal20}. We note that the differentiable population modeling approach taken here is naturally extensible to alternative characterizations of the halo profile that capture the splashback signature in the outer profile \citep[such as the fitting function in][]{diemer_kravtsov_2015}. This extension would proceed by using simulated merger trees to build a smooth model for the evolution of the splashback profile parameters of individual halos (as in \S\ref{sec:individual_conc}), and then by building a model for how the statistical distribution of the best-fitting parameters is connected to $\mpeak$ and the \dmah parameters (as in \S\ref{sec:conc_populations}). A directly analogous effort could similarly be applied to the recent dynamics-based model for halo profiles \citep{diemer22a}.

One promising application of our model would be to apply its predictions for modeling lensing, X-ray, and SZ effect profiles of galaxy groups and clusters using e.g., the Baryon Pasting (BP) model. The BP model is a physically motivated, computationally efficient approach for modeling X-ray and CMB skies \citep{shaw10,flender17,osato23}, for interpreting cross-correlation of weak lensing and SZ surveys \citep{osato18}, and for modeling upcoming multi-wavelength cross-correlation \citep{shirasaki20}. For example, it has recently been shown in \citet{green_etal20} that scatter in the thermal SZ effect is largely driven by variance in the assembly histories of cluster-mass halos, and the latter is faithfully captured by the \dmah approximation. Further motivation for this application of our model comes from recent advances that improve the sophistication of the observational measurement of cluster pressure profiles \citep[e.g.,][]{anbajagane_etal22_spt_shocks,keruzore_etal23_panco2}. These results create a timely opportunity for an effort to develop and calibrate a model for the {\it joint} dependence of the mean and scatter of these multi-wavelength cluster observables.

Another potentially interesting future application would be to adapt existing prediction pipelines for the evolution of halo substructure \citep[e.g.,][]{zentner_etal05,jiang_etal2021_satgen1} to leverage the \dmah and \dprof modeling ingredients. Recent progress in modeling subhalo orbital evolution has demonstrated the importance of accounting for the evolution of the host halo potential \citep{ogiya_etal21}, and so our model for $c(t)$ may be useful due to its capability to capture physically realistic correlations between halo assembly and internal structure \citep[see, e.g.,][]{jiang_vdb_2017}. In  future work, our larger aim is to unify theoretical predictions for the evolution of halo mass, internal structure, and substructure using the differentiable population modeling framework employed here.


\section{Conclusion}
\label{sec:conclusion}
Our principal findings are summarized as follows:
\ben
    \item We have built a new model for the evolution of the NFW concentration of individual dark matter halos across time. In our model, halo concentration evolves as a power-law function of time with a rolling index, and is characterized by three free parameters (see equation \ref{eqn:ind_model}). For halos identified by Rockstar in the Bolshoi-P and MDPL2 simulations with masses between $10^{11.5}M_{\sun}$ and $10^{15}M_{\sun}$, we have demonstrated that our model provides an unbiased approximation to the evolution of concentration for cosmic time $t\gtrsim2$~Gyr
    \item We have additionally built a model for the concentrations of halo populations, and demonstrated that our model  captures both {\it average} trends in the evolution of concentration, as well as its {\it diversity}, including physically realistic correlations with both halo mass and assembly history.
    \item Our python code, \href{https://github.com/BaryonPasters/diffprof}{\dprof}, is publicly available, can be installed with pip or conda, and includes Jupyter notebooks providing demonstrations of typical use cases. A parallelized script in the \dprof repository can be used to fit the concentration histories of individual simulated halos. The \dprof code also includes a convenience function that can be used to generate Monte Carlo realizations of the concentration histories of cosmologically realistic populations of halos. Precomputed fits for hundreds of thousands of halos in the Bolshoi-P and MDPL2 simulations are available at \url{https://portal.nersc.gov/project/hacc/aphearin/diffprof_data/}.

\een

\section*{Acknowledgements}
We thank Alex Alarcon, Matt Becker, Erwin Lau, Luisa Lucie-Smith, Phil Mansfield, Ismael Mendoza, and Kuan Wang for helpful discussions. The authors gratefully acknowledge the Gauss Centre for Supercomputing e.V. (www.gauss-centre.eu) and the Partnership for Advanced Supercomputing in Europe (PRACE, www.prace-ri.eu) for funding the MultiDark simulation project by providing computing time on the GCS Supercomputer SuperMUC at Leibniz Supercomputing Centre (LRZ, www.lrz.de). The Bolshoi simulations have been performed within the Bolshoi project of the University of California High-Performance AstroComputing Center (UC-HiPACC) and were run at the NASA Ames Research Center.
We acknowledge use of the Bebop cluster in the Laboratory Computing Resource Center at Argonne National Laboratory. Work done at Argonne was supported under the DOE contract DE-AC02-06CH11357. This work was supported in part by Yale Summer Experience Award and the DOE contract DE-AC02-06CH11357. DN is supported by NSF (AST-2206055) and NASA (80NSSC22K0821 \& TM3-24007X) grants. This work was performed in part at the Aspen Center for Physics, which is supported by National Science Foundation grant PHY-1607611. This work made extensive use of {\tt NumPy} \citep{numpy_ndarray}, {\tt SciPy} \citep{scipy}, Jupyter \citep{jupyter}, IPython \citep{ipython}, scikit-learn \citep{scikit_learn}, JAX \citep{jax2018github}, conda-forge \citep{conda_forge_community_2015_4774216}, Matplotlib \citep{matplotlib}, as well as
the Astrophysics Data Service (ADS) and {\tt arXiv} preprint repository.


\section*{Data Availability}
Software and data underlying this article is publicly available at the \dprof code repository on github, \url{https://github.com/BaryonPasters/diffprof}.

\bibliographystyle{mnras}
\bibliography{biblio}

\appendix

\renewcommand{\thefigure}{A\arabic{figure}}
\section{Fitting the Concentration Histories of Individual Halos}
\label{appendix:individual_conc}

In this appendix, we give a detailed description of our parametric model for the concentration history of individual halos, and how we fit the parameters of this model to the concentration history of simulated halos. As outlined in \S\ref{sec:individual_conc}, our model for the concentration history of individual halos is based on Eq.~\ref{eqn:ind_model}, which describes a power-law scaling between peak concentration and cosmic time, $c(t) = \cmin10^{\beta(t)},$ where the smoothly rolling power-law index $\beta(t)$ is controlled by a sigmoid function defined in Eq.~\ref{eqn:sigmoid}. Thus there are a total of four numbers that fully characterize the model: the early-time concentration, $\cmin,$ the late-time power-law index, $\cbl,$ the transition time parameter, $\tauc,$ and the transition speed parameter, $k.$ All results in the paper hold $k$ fixed to 4; to determine this particular value, we ran the optimization algorithm described below while allowing $k$ to be a free parameter, and then observed no appreciable changes in the quality of the fits when holding $k$ fixed to any value in its typical best-fitting range, $2\lesssim k\lesssim 5.$ Thus our model for the concentration evolution of individual halos has a total of 3 free parameters: $\cmin, \tauc,$ and $\cbl.$

In order to identify an optimal set of parameters for each halo, we searched our three-dimensional model parameter space, ${\theta},$ for the combination of $\cmin, \tauc,$ and $\cbl$ that minimizes the quantity $\mathcal{L}_{\rm MSE},$ defined as
\beq
\label{eq:mse_loss}
\mathcal{L}_{\rm MSE}(\theta)\equiv\frac{1}{N}\sum_{i}\left(w(t_i; {\theta})-v(t_i)\right)^2,
\eeq
where $w$ and $v$ are the base-10 logarithm of the predicted and simulated values of ${\rm c}(t)$ evaluated at a set of $N$ control points, $t_i.$ For the control points of each halo, we use the collection of snapshots in the halo's merger tree after a time $t_{\rm min},$ defined as
\beq
\label{eq:tminfit}
t_{\rm min}\equiv{\rm max}(t_{\rm cut},\ t_{\rm thresh}),
\eeq
where we set $t_{\rm cut}=2$~Gyr, and the quantity $t_{\rm thresh}$ is the first snapshot where the halo mass falls below $M_{\rm thresh};$ for the MDPL2 simulation, we use $M_{\rm thresh}=10^{11.25}\msun,$ and for the BPL simulation we use $M_{\rm thresh}=10^{10}\msun.$

\begin{figure*}
    \centering
    \includegraphics[width=12cm]{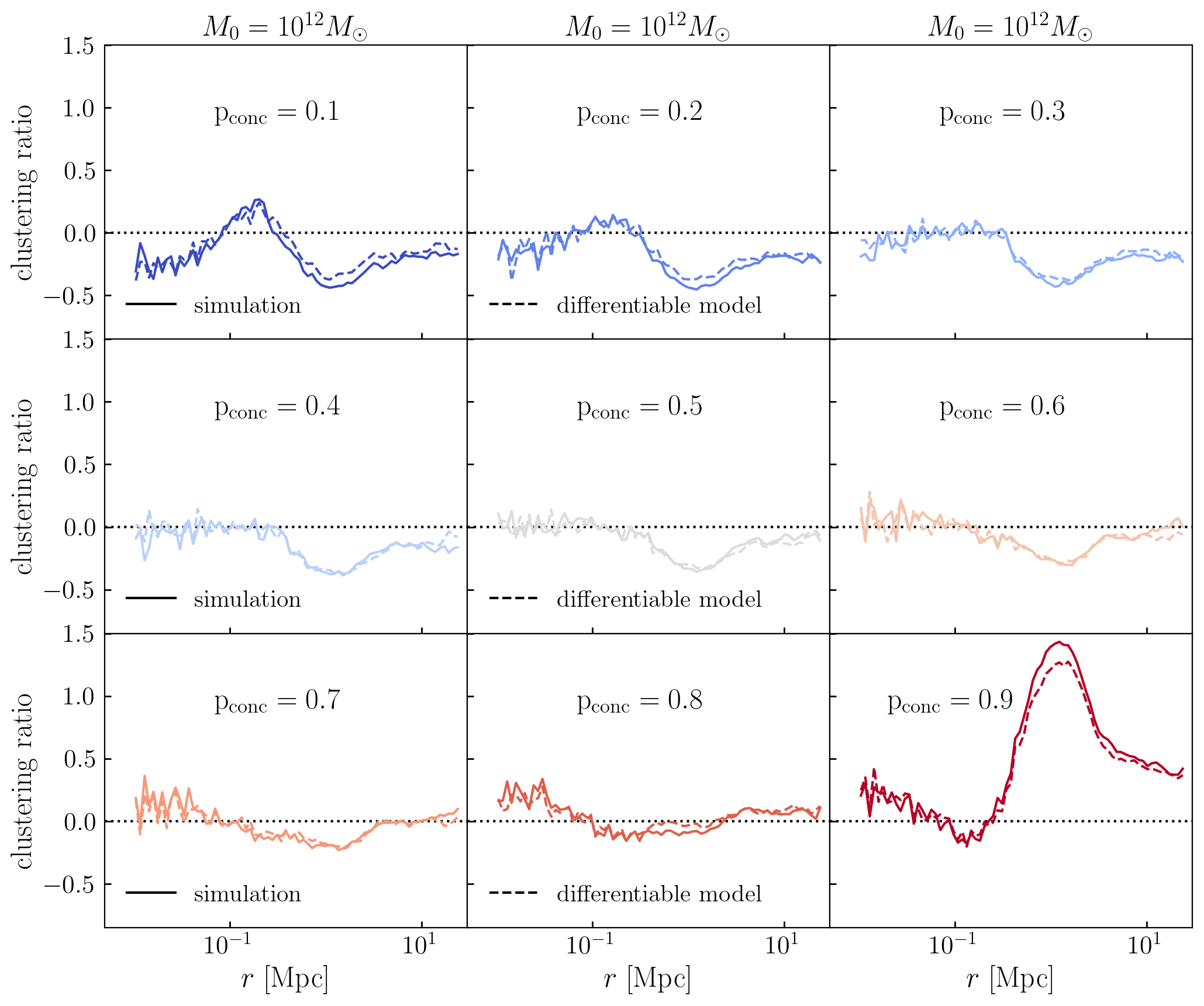}
    \caption{{\bf Concentration-dependence of halo clustering.} Same as Figure~\ref{fig:assembly_bias}, but using finer bins of concentration. Beginning with a sample of halos in the Bolshoi-P simulation with mass $10^{12}{\rm M_{\odot}},$ we divide the sample into nine bins of concentration. For each subsample, we compute $\xi_{\rm h\delta}(r),$ the cross-correlation between halos and dark matter particles. Solid curves show results for the case where concentration is taken directly from each halo’s simulated merger tree; dashed curves show results for the case where concentration is defined by the \dprof approximation. The figure provides further support for the conclusion that the correlation between halo concentration and the density field is retained when simulated concentration histories are approximated with \dprof.}
    \label{fig:assembly_bias_percentiles}
\end{figure*}

To minimize $\mathcal{L}_{\rm MSE}$ for each halo, we use either the {\tt scipy} implementation of the L-BFGS-B algorithm \citep{broyden_1970_B_in_BFGS,fletcher_1970_F_in_BFGS,goldfarb_1970_G_in_BFGS,shanno_1970_S_in_BFGS}, or the JAX implementation of the Adam algorithm  \citep{kingma_ba_adam_2015}. Both of these algorithms require calculating the gradients $\partial\mathcal{L}_{\rm MSE}/\partial\theta,$ which is straightforward using  automatic differentiation due to the JAX-based implementation of our model. In the small number of cases where the L-BFGS-B fitter fails to converge, we fall back on Adam, taking 300 adaptive steps using step-size parameter 0.001.

When minimizing $\mathcal{L}_{\rm MSE}$, the parameters we actually vary in the gradient descent are $\cminprime,\, \cblprime,$ and $\taucprime,$ each of which is an unbounded version of its counterpart that appears Eq.~\ref{eqn:ind_model}, with associated transformations defined as follows:
\beq
\label{eq:param_bound_transformation}
\cmin & = & \mathcal{S}(\cminprime\vert k_{\rm b}, x_{\rm b}, L_{\cmin}, U_{\cmin})\nonumber\\
\cbl & = & \mathcal{S}(\cblprime\vert k_{\rm b}, x_{\rm b}, \cmin, U_{\cbl})\\
\tauc & = & \mathcal{S}(\taucprime\vert k_{\rm b}, x_{\rm b}, L_{\tauc}, U_{\tauc})\nonumber
\eeq
The quantities $k_{\rm b}=0.1$ and $x_{\rm b}=0$ are common to each bounding function in Eq.~\ref{eq:param_bound_transformation}. We set $L_{\cmin}=2,$ and $U_{\cmin}=5.5,$ restricting the fitter to seek relatively small values of the early-time concentration. We set $L_{\tauc}=1,$ $U_{\tauc}=30,$ and $U_{\cbl}=300$. The lower bound for $\cbl$ is the early-time value $\cmin$, so that our optimization algorithm only seeks best-fitting solutions $c(t)$ that are strictly non-decreasing, with a floor given by $\cmin.$

Both the functional form defined by Eq.~\ref{eqn:ind_model}, as well as these bounds, essentially impose a {\it narrative} on how concentration evolves over time. The narrative we impose is essentially the same physical picture reviewed in \S\ref{sec:intro}: in the fast-accretion regime, concentration maintains a small, constant value, and increases at later times as halo growth slows down. Additionally, in the narrative of this model, concentration evolves in entirely smooth fashion, and the diversity of concentrations manifests in the diversity of smooth trajectories captured by the statistical distribution of $\cmin, \tauc,$ and $\cbl.$ 

One can imagine alternatives to approximating concentration history that are less restrictive and thereby capture additional complexities, such as the transient fluctuations associated with merger events that are visible in Figure~\ref{fig:individual_conc_fit}. For example, models based on machine learning have capability to capture such fluctuations by offering essentially maximal flexibility, and by allowing the simulated dataset to fully ``speak for itself". With the \dprof model, we have taken the complementary approach of seeking a {\it minimally} flexible model that {\it i)} embodies the physical interpretation of the classical picture of concentration evolution, and {\it ii)} accurately captures the broad characteristics of each individual halo's evolution. These two approaches have potential to work in concert, with a \dprof type model describing the smooth evolutionary component of concentration, and a machine learning model capturing the correlated residuals of the fits. While beyond the scope of the present paper, we intend to explore such a hybrid approach in future work.

Even though our model imposes these rather stringent expectations on the evolution of the concentration evolution of individual halos, the results in the main body of the text demonstrate that the \dprof model is able to give an unbiased approximation of ${\rm c}(t)$ with reasonably low scatter. Additionally in Figure~\ref{fig:assembly_bias}  we showed that residual errors in our fits are nearly uncorrelated with the density field. Figure~\ref{fig:assembly_bias_percentiles} provides a finer-grained analysis of this finding. We remind the reader that in generating Figure~\ref{fig:assembly_bias}, we started with sample of halos in the BPL simulation with mass $\mzero=10^{12}\msun,$ split the sample in half according to $\pc,$ and for each subsample we calculated $\xi_{\rm h\delta}(r),$ the two-point cross function between the halo centers and particle tracers of the dark matter density field. We repeat this analysis in Figure~\ref{fig:assembly_bias_percentiles}, only rather than splitting the halo sample merely in half, instead we split it into $9$ different subsamples.

Each panel of Figure~\ref{fig:assembly_bias_percentiles} shows the fractional difference in $\xi_{\rm h\delta}(r)$ between the subsample and the entire sample of halos of this mass, with the value of $\pc$ of the subsample indicated by the in-panel annotations. The collection of panels supports the same basic conclusion as drawn in the main body of the text: the residual errors of the \dprof model are largely uncorrelated with the density field. The discrepancy between the solid and dashed curves is largest for the highest-concentration subpopulation shown in the bottom right panel. This particular outlier population of halos is likely made up of a significant fraction of ``splashback" subhalos, and so it is unclear to what extent this discrepancy would be reduced if our simulation data were based on alternative halo boundary definitions; see \S\ref{subsub:outlier_halos} for further discussion.

We conclude this appendix by noting that our model for the concentrations of individual halos does not hard-wire any particular relationship between ${\rm c}(t)$ and $\mhalo(t),$ as neither the parameters of our model, nor their bounds, have any mathematical connection to the evolution of halo mass. This distinguishes our model for concentration growth from other models in the literature, which {\it directly} connect the concentration history of individual halos to some particular measure of halo formation time \citep[e.g.,][]{Zhao_2009,ludlow_etal13,correa_etal15}. In our approach, the connection between concentration and halo mass assembly history is expressed in terms of correlations between $\mhalo(t)$ and $\cmin, \tauc,$ and $\cbl$ that emerge in the distributions of best-fitting values. We capture this connection in our model for the concentration history of halo populations using the techniques detailed in the remaining appendices.

\renewcommand{\thefigure}{B\arabic{figure}}
\section{Differentiable predictions of a probabilistic model}
\label{appendix:differentiable_populations}

In this section, we outline the general techniques we use when optimizing the parameters of a model whose fundamental predictions are probability distributions, i.e., a probabilistic model. For pedagogical purposes, here we will discuss a low-dimensional toy model that is simpler than \dprofpop, but has the same hierarchical structure. Readers already familiar with autodiff and probabilistic programming may wish to skip directly to Appendix \ref{appendix:diffprofpop_definition} for specifics on the \dprofpop model. 

Consider a two-level hierarchical model in which knowledge of hyper-parameters $\Psi$ allows one to make predictions for $P_{\Psi}(\theta),$ a statistical distribution of parameters $\theta.$ For simplicity, let's consider $P_{\Psi}(\theta)$ to be a multi-variate Gaussian distribution of a two-dimensional parameter space, $\theta=\{x, y\}.$ Thus the hyper-parameters $\Psi$ control the mean value $\mu=\{\bar{x}, \bar{y}\},$ and the covariance matrix $\Sigma[x, y].$ In this toy model, we are equipped with some prediction function $Q(x, y),$ so that for any particular values of the parameters $\theta=\{x, y\},$ we are able to calculate the quantity $Q.$ As we will see in the next section, this general form of a hierarchical model maps directly onto the problem of modeling the cosmological abundance of halo concentrations, since in that case the \dprof parameters $\theta_{\rm NFW}$ supply a prediction function for $Q=c(t),$ and the \dprofpop parameters specify a statistical distribution $P(\theta_{\rm NFW})$ that enables a prediction for $P(c(t)).$

Our goal in the optimization of the hyper-parameters $\Psi$ is to make a prediction for some target distribution, $P_{\rm target}(Q).$ For the sake of specificity, we will use the mean $\langle Q\rangle$ and variance $\sigma^2(Q)$ of the target distribution to construct a summary statistic that quantifies the goodness of fit. The equations below show the general form of the convolutional integrals that must be calculated in order to make a model prediction for the mean and variance that can be compared to the targets:
\beq
\label{eq:generic_sumstats}
\langle Q\vert\Psi\rangle \equiv \int\dd x\dd y P_{\Psi}(x, y)\cdot Q(x, y)\\
\label{eq:generic_sumstats2}
\sigma^2(Q\vert\Psi) \equiv \int\dd x\dd yP_{\Psi}(x, y)\cdot \left(Q-\langle Q\rangle\right)^2
\eeq

There are numerous computational techniques that can be used to calculate the convolutional integrals appearing in Eqs.~\ref{eq:generic_sumstats}\&\ref{eq:generic_sumstats2}. So long as the integrands are reasonably smooth, then classical iterative integration algorithms \citep[e.g.,][]{romberg_integration_1955} have no trouble quickly converging to a high-accuracy result. Alternatively, since for any value $\Psi$ the distribution $P_{\Psi}(x,y)$ is just a two-dimensional Gaussian, then it is also straightforward to adopt a stochastic Monte Carlo approach in which one draws random realizations of $P_{\Psi}(x, y),$ and then simply computes the average and variance of the values $Q(x, y)$ from the generated distribution. Either of these methods could be implemented based on widely-used scientific computing libraries such as {\tt scipy} \citep{scipy}. 

Instead, we carry out these integrals by taking the PDF-weighted average of $Q(x, y)$ for every point in $\mathcal{G},$ a {\it uniform} grid in $\{x, y\}$ that spans the support of the PDF. For example, to predict the mean value of $Q$ for a point $\Psi$ in the hyper-parameter space:
\beq
\label{eq:toy_pdf_first_moment}
\langle Q\vert\Psi\rangle = \frac{\sum_{\mathcal{G}} Q(x_i, y_i)\cdot P_{\Psi}(x_i, y_i)}{\sum_{\mathcal{G}} P_{\Psi}(x_i, y_i)}.
\eeq
In this two-dimensional toy problem, using a simple, linearly-spaced grid in each dimension is entirely practical for $\mathcal{G},$ but one could alternatively calculate an unbiased result using grid-points $\{x, y\}$ with a uniform random distribution, or with a quasi-random space-filling distribution such as a Latin Hypercube \citep[LH,][]{mckay_etal79_latin_hypercube} or Sobol Sequence \citep{sobol_sequence_1967}. The main advantage of a space-filling design is that the result converges more quickly as the number of points in the grid increases. In optimizing \dprofpop, we use an LH-based design in anticipation of extending the \dprof model in follow-up work that jointly incorporates additional halo and galaxy properties. While such downstream models will necessarily increase the dimension of these PDF-convolution integrals, the dimensionality will likely remain far smaller than present-day data science applications using the same methods to carry out such integrals in hundreds of dimensions \citep{dick_kuo_sloan_2013}.

\begin{figure}
    \centering
    \includegraphics[width=8cm]{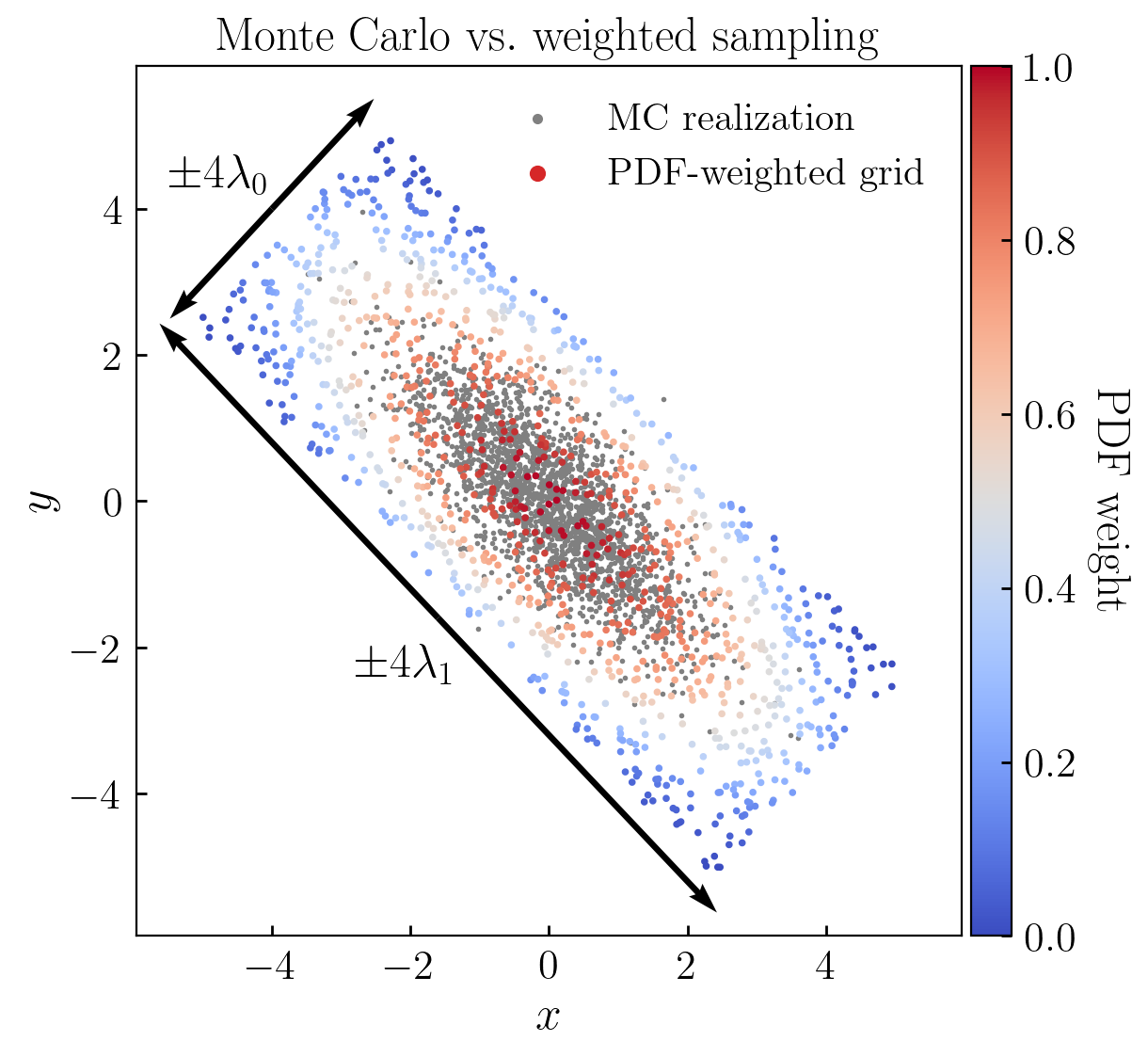}
    \caption{{\bf Visual demonstration of PDF-weighted sampling.} The figure shows a scatter plot of a toy model used to demonstrate the equivalence of Monte Carlo vs. weighted sampling. The axes show the two-dimensional parameter space, $\theta=\{x, y\},$ which is distributed according to a multivariate Gaussian. The hyper-parameters $\Psi$ control the mean and covariance of this multivariate Gaussian, $P_{\Psi}(x, y).$ Gray points show a random Monte Carlo realization of the PDF, so that the density of points reflects the concentration of the PDF towards its center. Colored points are {\it uniformly} distributed in $\theta$-space according to a Latin Hypercube, and are color-coded by the value of the PDF. As described in Appendix~\S\ref{appendix:differentiable_populations}, either method of sampling can be used to make predictions for the concentrations of halo populations, but the PDF-weighted grid method admits and more natural and GPU-efficient implementation in an autodiff library such as JAX. See text for details.}
    \label{fig:weighted_sampling}
\end{figure}

Figure~\ref{fig:weighted_sampling} gives a visual illustration of the weighted-sampling method we adopt to calculate results of the PDF convolutions in Eqs.~\ref{eq:generic_sumstats}-\ref{eq:generic_sumstats2}. The figure shows a two-dimensional scatter plot of a Gaussian-distributed variable $\theta=\{x, y\}.$ The gray points show a random Monte Carlo realization of a distribution centered at the origin with covariance ${\Sigma}[x, y]=[\ [1, -c ],\ [-c, 1]\ ],$ where $c=3/4.$ The colored points are distributed according to a Latin Hypercube quasi-random grid, $\mathcal{G};$ the axes of $\mathcal{G}$ are aligned with the eigenvectors of $\Sigma,$ and the side lengths span $\pm4\lambda_i,$ where $\lambda_i$ is the $i^{\rm th}$ eigenvalue of $\Sigma.$ The grid points are color-coded according to the value of the Gaussian PDF, so that points in the centroid of the distribution are up-weighted relative to points in the outskirts.

Calculating Eq.~\ref{eq:toy_pdf_first_moment} using the PDF-weighted grid method above simplifies the propagation of gradients of parameters $\Psi$ through model predictions for summary statistics of Q. For example:
\beq
\label{eq:toy_pdf_convolution}
\partial \langle Q\rangle/\partial\Psi = \sum_{\mathcal{G}}Q(x_i, y_i)\cdot\frac{\partial}{\partial\Psi}P_{\Psi}(x_i, y_i).
\eeq
Provided that the parametrized model $P_{\Psi}(x, y)$ admits an implementation in an autodiff library, then calculating derivatives of $\langle Q\rangle$ with respect to $\Psi$ is straightforward. This approach also makes it easy to pre-generate a uniform grid in advance, eliminating stochasticity from all downstream gradient computations; this also simplifies the declaration of memory resources in the software implementation, and allows libraries such as JAX to more effectively vectorize the computation on GPUs and other accelerator devices.

\renewcommand{\thefigure}{C\arabic{figure}}
\section{Modeling the Concentration Histories of Halo Populations}
\label{appendix:diffprofpop_definition}

In this appendix, we describe the specific form of \dprofpop, our model for the cosmological abundance of halo concentrations. The goal of \dprofpop is to capture $P(c(t)\vert M_0, \tform),$ and as outlined in \S\ref{subsec:diffpop}, the basis of our approach is  the \dprof model, which approximates the evolution of the NFW concentration of an individual halo with parameters $\theta_{\rm NFW}=\{\cmin, \tauc, \cbl\}.$ The \dprofpop model defines $P(\theta_{\rm NFW}\vert M_0, \pc),$ the statistical distribution of $\theta_{\rm NFW},$ including dependence on present-day halo mass, $M_0,$ and the percentile of halo formation time, $\pc\equiv P(<\tperc{50}\vert M_0).$ Since $\theta_{\rm NFW}$ defines an analytical and deterministic expression for $c(t),$ then the cosmological abundance of halo concentrations is fully specified once $P(\theta_{\rm NFW}\vert M_0, \pc)$ is characterized.

In defining the statistical distribution of $\theta_{\rm NFW},$ the quantity we characterize in practice is $P(\tilde{\theta}_{\rm NFW}\vert M_0, \pc),$ i.e., the PDF of the {\it unbounded} version of $\theta_{\rm NFW},$ where the $\theta_{\rm NFW}-\tilde{\theta}_{\rm NFW}$ relationship between is given by Eq.~\ref{eq:param_bound_transformation}. This modeling choice guarantees that all generated trajectories of $c(t)$ will respect the physical constraints on $\cmin, \tauc,$ and $\cbl$ that we mentioned in \S\ref{sec:individual_conc} and detailed in Appendix~\ref{appendix:individual_conc}.

We model $P(\tilde{\theta}_{\rm NFW})$ as the product of two Gaussians:
\beq
\label{eq:dpp_separation}
P(\tilde{\theta}_{\rm NFW}\vert\mzero,\pc) = P(\cminprime\vert\mzero)\cdot P(\taucprime, \cblprime \vert \mzero, \pc).
\eeq
In Eq.~\ref{eq:dpp_separation}, $P(\cminprime)$ is a one-dimensional Gaussian distribution with mean $\mu_{\cminprime}$ and variance $\Sigma_{\cminprime}\equiv\sigma^2(\cminprime)$ that depend only upon $\mzero,$ while the distribution $P(\taucprime, \cblprime)$ is a two-dimensional Gaussian with mean $\mu_{\taucprime, \cblprime}$ and covariance matrix $\Sigma_{\taucprime, \cblprime}$ that depends jointly upon both $\mzero$ and $\pc.$ All dependencies upon halo mass are defined in terms of $\lgmzero\equiv\lgg\mzero.$ In the remainder of this appendix, we merely describe the mathematical forms of these four parametrized functions, relegating discussion of our determination of the specific values of the parameters to Appendix~\ref{appendix:population_conc_fitting}.

We model the functions $\mu_{\cminprime}(\lgmzero)$ and $\Sigma_{\cminprime}(\lgmzero)$ according to a simple sigmoid function, $\mathcal{S}(x),$ (see Eq.~\ref{eqn:sigmoid}), where $x=\lgmzero.$ The definition we adopt for $\mathcal{S}(x)$ depends upon four parameters, but in each case we find that only two degrees of freedom are warranted by the available simulation data and our choice of target summary statistics.

The characterization of $\mu_{\taucprime}$ and $\mu_{\cblprime}$ is more involved because these two functions exhibit simultaneous dependence upon $\lgmzero$ and $\pc.$ First, at fixed halo mass, we characterize the $\pc$-dependence of $\mu_{\taucprime}$ and $\mu_{\cblprime}$ according to two independently-specified ``sigmoid-slope" functions, $\mathcal{R}(x),$ defined as follows:
\beq
\label{eq:sigslope}
\mathcal{R}(x) = y_0 + \mathcal{S}(x)\cdot(x-x_0),
\eeq
where $\mathcal{S}(x)$ is defined in Eq.~\ref{eqn:sigmoid}. The behavior of a sigmoid-slope function is more flexible than a sigmoid function because $\mathcal{R}(x)$ able to capture non-monotonic dependence upon $x$ since $\mathcal{S}(x)$ can take on both negative and positive values. Thus for both $\mu_{\taucprime}$ and $\mu_{\cblprime}$, the independent variable in Equation~\ref{eq:sigslope} is $x=\pc.$ Our sigmoid-slope function depends upon five parameters, and each of these parameters in principle could be permitted to depend upon $\lgmzero.$ In practice, we only allow three of these parameters to depend upon halo mass: $y_0$ and the two slope parameters $y_{\rm lo}$ and $y_{\rm hi}.$ We characterize this $\lgmzero$-dependence via a sigmoid function, e.g., $y_0(\lgmzero)$ varies as a function of halo mass according to its own independently-defined $\mathcal{S}(x).$ As described in Appendix~\ref{appendix:population_conc_fitting}, when programmatically varying $y_0(\lgmzero),$ $y_{\rm lo}(\lgmzero),$ and $y_{\rm hi}(\lgmzero),$ we find that only two of the available four sigmoid parameters are warranted by our target summary statistics and desired accuracy.

In order to model the two-dimensional covariance matrix $\Sigma_{\taucprime, \cblprime},$ we make use of the Cholesky matrix decomposition. Briefly, the Cholesky decomposition is a lower-triangular matrix, $L,$ whose relationship to the covariance matrix is given by $\Sigma=L\cdot L^{\rm T};$ provided that $\Sigma$ is a real-valued, symmetric, positive-definite matrix (satisfied by all non-singular covariance matrices), then $L$ is unique, real-valued, with strictly positive diagonal entries whose product is equal to the determinant of $\Sigma.$ We use JAX to calculate $L$ so that our decomposition will be differentiable, but many modern linear algebra libraries have efficient implementations of the Cholesky decomposition \citep[for a modern review, see][]{higham_2009_cholesky_review_article}.

In modeling $\Sigma_{\taucprime, \cblprime},$ the quantities we parameterize are the 3 entries of its associated Cholesky matrix:
\beq
\label{eq:cholesky}
L&\equiv&
  \begin{bmatrix}
   A & 0   \\
   C & B
   \end{bmatrix}.
\eeq
When fitting the \dprofpop model as described in Appendix~\ref{appendix:population_conc_fitting}, the quantities we parametrize are $a\equiv\lgg A$ and $b\equiv\lgg B$ to ensure that $\Sigma=L\cdot L^{\rm T}$ will always be positive definite, but the parameter $C$ can take on any value on the real line.

Our characterization of the $\lgmzero$- and $\pc$-dependence of $a, b,$ and $C$ mirrors our treatment of this dependence for $\mu_{\taucprime, \cblprime}.$ First, at fixed $\mzero,$ we parametrize $a, b$ and $C$ to depend upon $\pc$ according to a sigmoid-slope function, $\mathcal{R}(x).$ To capture the $\mzero$-dependence, for each variable we allow the best-fitting sigmoid-slope parameters to depend upon $\lgmzero.$ We give a detailed account of our fitting techniques in the next section.

\renewcommand{\thefigure}{D\arabic{figure}}
\section{Fitting the \dprofpop Parameters}
\label{appendix:population_conc_fitting}

In this section, we apply the techniques reviewed in Appendix~\ref{appendix:differentiable_populations} to optimize the parameters $\Psi_{\rm NFW}$ of the \dprofpop model defined in Appendix \ref{appendix:diffprofpop_definition}. Our goal for the calibration of $\Psi_{\rm NFW}$ is to reproduce a realistic diversity of individual, smooth trajectories of $c(t),$ such that the distribution $P(c(t)\vert M_0, \pc)$ seen in simulations is well-approximated by \dprofpop. We begin in \ref{subapp:dpp_targets} by describing how we arrive at the specific target data vectors, and conclude in \ref{subapp:dpp_opt} by detailing the definition of our specific loss function and its minimization.

\subsection{Target data}
\label{subapp:dpp_targets}

The target data for \dprofpop are defined by the merger trees in the Bolshoi-P and MDPL2 simulations for halos divided into bins of present day halo mass, $\mzero,$ and halo formation time percentile, $\pc\equiv P(<\tperc{50}\vert\mzero).$ To calculate $\pc,$ we use the \href{https://halotools.readthedocs.io/en/latest/api/halotools.utils.sliding_conditional_percentile.html}{sliding\_conditional\_percentile} function in halotools, which uses a sliding window to provide a bin-free estimate of the rank-order percentile of $\tperc{50}$ conditioned on halo mass. We select host halos in bin of $\mzero$ with width $0.1$dex, and a bin of $\pc$ with width 0.1, using the Bolshoi-P simulation for $\mzero\leq10^{13.5}\msun,$ and MDPL2 for larger halo masses. Our target means and variances are defined at particular values of cosmic time, $t,$ measured in Gyr; we use every simulation snapshot available across the range $2<t<13.8,$ and estimate the mean and variance of $\lgc$ amongst the halos in the bin. In detail, we use the trimmed mean, $\mu_{\rm q},$ and trimmed variance, $\sigma_{\rm q},$ calculated using {\tt scipy.stats.mstats}; within {\tt scipy}, these quantities are computed after first excluding the objects lying outside the outer $q=10\%$ quantile range. Our target summary statistics are formulated in terms of $\lgc$ because the distribution of $c(t)$ is close to log-normal, and so logarithmic averages correspond more closely to the median.

Our goal with \dprofpop is to faithfully recover the diversity of {\it smooth} concentration histories for a population of halos. As our  model for $c(t)$ of individual halos does not account for transient fluctuations associated with mergers, for our target data we use each individual halo's best-fitting $c(t)$ to define the target mean and variance of $P(c(t)\vert\mzero,\pc).$ If we were to instead use the directly simulated histories to define the target one-point functions, then our model would instead converge to a result with unrealistically large variance in the smooth trajectories traced by real halos. Using the directly simulated values as target data would require building an additional model component describing the contribution to the variance from mergers, which is beyond our present scope and is the subject of future work (see \S\ref{sec:discussion} for further discussion).

\subsection{Loss function minimization}
\label{subapp:dpp_opt}

Our loss function is defined in terms of three separate predictions:
\begin{align}
\label{eq:sumstats}
\langle Q\vert\mzero,\pc\rangle &\equiv \int\dd\thetanfw P(\thetanfw\vert\mzero,\pc)\cdot Q\\
\langle Q\vert\mzero\rangle &\equiv \int\dd\thetanfw\dd\pc P(\thetanfw\vert\mzero,\pc)\cdot Q\nonumber\\
\sigma^2(Q\vert\mzero) &\equiv \int\dd\thetanfw\dd\pc P(\thetanfw\vert\mzero,\pc)\cdot \left(Q-\bar{Q}\right)^2\nonumber
\end{align}
We use Eq.~\ref{eq:mse_loss} to calculate the loss function associated with each of these three predictions separately, and then sum the three results to compute the net loss. 

The predictions of \dprofpop are controlled by the parametrized behavior of the functions defined in Appendix~\ref{appendix:diffprofpop_definition}. Optimizing the parameters controlling these functions requires an initial guess for the parameters. In order to determine this initial guess, we first directly inspect the $\pc$-dependence of the distribution of best-fitting \dprof parameters in each target bin, $\mzero.$ Within each halo mass bin, we coarsely tune the parameters of each sigmoid- and sigmoid-slope function so that $\mu_{\cminprime}(\lgmzero),$ $\mu_{\taucprime}(\lgmzero)$ and $\mu_{\cblprime}(\lgmzero)$ pass through the visual centroid of the distributions. We similarly use visual inspection to approximate the rough bounds on each of the parameters. We proceed in this fashion, bin-by-bin, until we have obtained an initial rough estimate of $\Psi^0_{\rm NFW}$, the full set of parameters that govern the predictions of \dprofpop.

In the next phase of model optimization, we use the Nelder-Mead algorithm \citep{nelder_mead_optimization_1965} implemented in {\tt scipy} in order to further refine our estimate for $\Psi_{\rm NFW}.$ Because Nelder-Mead is a direct-search algorithm based on a simplex decomposition of the parameter space, no gradients are required in this phase of the optimization, and so here we relied on a stochastic Monte Carlo implementation of \dprofpop in order to begin at the point $\Psi^0_{\rm NFW}$ and terminate at a new estimate, $\Psi^1_{\rm NFW}.$

Using the procedure described above to define the initial guess, $\Psi^1_{\rm NFW},$ we use the Adam algorithm \citep{kingma_ba_adam_2015} to define the adaptive learning rate we use in our final gradient descent. In searching for the best-fitting point, we do not allow the parameters to vary arbitrarily, but rather we only permit the gradient descent algorithm to search within a relatively narrow range of the initial guess. We carry out this restricted search by using the same sigmoid bounding technique described in Appendix~\ref{appendix:individual_conc}, only here we use bounds on the parameters $\Psi_{\rm NFW}$ via the visual inspection described at the beginning of this section. We refer the reader to the {\tt bpl\_dpp.py} module of the \dprof code for the specific best-fitting values that result from this optimization exercise.


\bsp	
\label{lastpage}
\end{document}